\documentclass[fleqn,usenatbib]{mnras}

\usepackage{newtxtext,newtxmath,subfigure}
\usepackage{orcidlink}

\usepackage[T1]{fontenc}

\DeclareRobustCommand{\VAN}[3]{#2}
\let\VANthebibliography\thebibliography
\def\thebibliography{\DeclareRobustCommand{\VAN}[3]{##3}\VANthebibliography}

\usepackage{graphicx}	
\usepackage{amsmath}	

\makeatletter

\makeatletter

\makeatletter
\DeclareRobustCommand{\HI}{%
  \mbox{H\check@mathfonts\fontsize\sf@size\z@\selectfont I}%
}
\makeatother
\makeatletter
\DeclareRobustCommand{\HII}{%
  \mbox{H\check@mathfonts\fontsize\sf@size\z@\selectfont II}%
}
\makeatother

\title[Kinematic asymmetries in cluster galaxies]{Cluster passage driving galaxy kinematic and structural evolution in the SAMI Galaxy Survey}

\author[R. S. Bagge]{
R. S. Bagge \orcidlink{0009-0002-2753-3248},$^{1,2}$\thanks{E-mail: r.bagge@unsw.edu.au},
C. Foster \orcidlink{0000-0003-0247-1204},$^{1,2}$
S. Brough,$^{1,2}$
O. Çakır,$^{3,4,2}$
L. Cortese,$^{5,2}$
F. D'Eugenio,$^{5,6}$
S. Croom \orcidlink{0000-0003-2880-9197},$^{6,2}$
\newauthor
M. Owers,$^{3,4,2}$
J. van de Sande$^{1,2}$
\\
$^{1}$School of Physics, University of New South Wales, Kensington, NSW, 2032, Australia\\
$^{2}$ARC Centre of Excellence for All Sky Astrophysics in 3 Dimensions (ASTRO 3D)\\
$^{3}$School of Mathematical and Physical Sciences, Macquarie University, NSW 2109, Australia,\\
$^{4}$Astrophysics and Space Technologies Research Centre, Macquarie University, Balaclava Road, Sydney, NSW 2109, Australia,\\
$^{5}$Kavli Institute for Cosmology, University of Cambridge, Madlingley Rd, Cambridge, CB3 OHA, UK,\\
${6}$Cavendish Laboratory - Astrophysics Group, University of Cambridge, 19 JJ Thomson Avenue, Cambridge CB3 0HE, UK\\
$^{7}$ICRAR,The University of Western Australia, 7 Fairway, Crawley, WA, 6009, Australia\\
$^{8}$Sydney Institue for Astronomy, School of Physics, University of Sydney, NSW 2006, Australia\\
}

\date{Accepted XXX. Received YYY; in original form ZZZ}

\pubyear{2023}

\begin{document}
\label{firstpage}
\pagerange{\pageref{firstpage}--\pageref{lastpage}}
\maketitle

\begin{abstract}
The cluster environment can have a significant impact on galaxy evolution. We study the impact that passage through a cluster has on stellar and ionised gas kinematics for galaxies within the Sydney-AAO Multi Integral field (SAMI) Galaxy Survey. We compute the kinematic asymmetry $v_{\rm asym}$ in the line-of-sight stellar and ionsied gas velocity maps to quantify how the cluster environment disturbs the kinematics of the stars and ionised gas. We find a significantly higher fraction of galaxies with elevated gas asymmetries in clusters compared to non-cluster environments (17$^{+2}_{-3}$\%, 26/154 vs. 11$^{+1}_{-1}$\%, 72/751), with these galaxies most likely being recent infallers passage based on their position in projected-phase-space. Compared to cluster galaxies without elevated gas asymmetries, cluster galaxies with elevated gas asymmetries have, on average, more centrally concentrated star-formation. Finally, we find the highest fraction of galaxies with elevated gas asymmetries in clusters likely to host significant substructure or be dynamically complex. Our findings are consistent with the scenario of galaxies falling into clusters, either individually or in groups, and undergoing disk-fading and a redistribution of gas, due to ram pressure stripping experienced during pericentre passage.
\end{abstract}




\section{Introduction}
Galaxy clusters are the highest-density environments that individual galaxies can reside in, and such an extreme environment can drastically impact their evolution. It is a well established fact that the number of Early-Type Galaxies (ETGs) increases as the density of the environment increases, while the number of Late-Type Galaxies (LTGs) decreases \citep[morphology-density relation; e.g.,][]{1997ApJ...490..577D,2005MNRAS.356.1155C,2012MNRAS.423.3478H}. Similar to this morphology-density relation, the host environment of a galaxy can impact the Star Formation Rate (SFR), with the number of star-forming galaxies decreasing as the environmental density increases in favour of quenched (non-star-forming) galaxies \citep[e.g.,][]{2002MNRAS.334..673L,2009MNRAS.400.1225C,2010ApJ...721..193P}, the above trends have been attributed to what is now known as `environment quenching'. 

Environmental quenching can manifest in different ways depending on the environment the galaxy inhabits and whether it dominates that environment (i.e., central vs. satellite, cluster or group). Central galaxies (those that are the most massive within their respective environments) typically quench through internal, secular processes \citep[e.g., feedback from star-formation (SF) and Active Galactic Nuclei, exhaustion of fuel for star-formation;][]{1980ApJ...237..692L,2002ApJ...577..651B,2006MNRAS.365...11C,2007ApJ...664..204S}, and to a lesser extent their environment \citep{2020MNRAS.492...96B,2024MNRAS.530.5072M}. Conversely, satellite galaxies (those that do not dominate their environment by mass) are more susceptible to environmental quenching mechanisms, which can manifest as cessation of gas accretion or stripping of gas within the galaxy. For satellite galaxies in a cluster, environmental quenching acts primarily through hydrodynamical interactions between the hot (T>10$^7$ K) Intracluster Medium (ICM) and the cold gas within the satellite galaxy. As satellites move through the ICM they experience ram pressure stripping \citep[RPS;][]{1972ApJ...176....1G}, removing the less gravitionally bound gas in the outskirts of galaxies. Similarly, the cold gas can be removed through evaporation as it interacts with the ICM \citep{1977Natur.266..501C} or be viscously stripped \citep{1982MNRAS.198.1007N}. Environmental quenching is not restricted to hydrodynamic processes alone. Due to the deep gravitational potential of the cluster, cluster members can have large relative velocities ($\sim$500 kms$^{-1}$) causing a number of high-speed interactions which can happen multiple times as galaxies pass through the cluster \citep{1996Natur.379..613M,2015MNRAS.452..616D,2017ApJ...844...59B} or be tidally stripped as they pass close to the centre of the cluster \citep{1990ApJ...350...89B,1999ApJ...510L..15B}. These processes (can) work in tandem to effectively quench satellite galaxies \citep[see][ for recent reviews]{2021PASA...38...35C,2022A&ARv..30....3B}.

As well as quenching, there is a strong correlation between the kinematic structure of a galaxy and its environment. Early Integral Field Spectroscopic (IFS) survey, ATLAS$^{\rm 3D}$ examined the `kinematic morphology-density' relation in the Virgo Cluster \citep{2011MNRAS.416.1680C}, where the authors found that `slow-rotating' ETGs will preferentially be found in cores of big clusters. Later IFS studies using the Sydney-AAO Multi Integral field (SAMI) Galaxy Survey \citep{2015MNRAS.447.2857B} and Mapping Nearby Galaxies at APO (MaNGA) Survey \citep{2015ApJ...798....7B} explored this further and found a weak relation between `slow-rotator' fraction and host cluster mass, $M_{200}$, but found there was a trend between local overdensity, commonly parameterized as distance to 5$^{\rm th}$ nearest neighbour, $\Sigma_5$, with slow-rotators typically found in the substructure within the cluster \citep{2014MNRAS.443..485F,2017ApJ...844...59B,2017ApJ...851L..33G}. However, the connection between environment and stellar kinematics is still nuanced. While correlations between slow-rotator fraction, mass and environment exist, and act independently \cite[e.g.,][]{2021MNRAS.508.2307V}; after accounting for these correlations, it appears mean stellar age is the primary parameter determining whether a galaxy is a slow or fast rotator, with no residual correlation with environment for low mass galaxies \citep[e.g., probed with either $\Sigma_5$ or halo mass;][]{2024MNRAS.529.3446C}. This would suggest the main determinant of galaxy spin is either the physical conditions of initial star-formation (e.g., stars being born dynamically hotter), or when the galaxy was quenched. This lack of a residual environmental correlation once accounting for stellar age persists to $z\sim0.3$, offering more evidence for the secondary role environment has on the dynamical evolution over the last 4 Gyrs \citep{2025arXiv250108461F}.

Disentangling the cumulative effects of environment on galaxy evolution can be difficult since the different physical processes can work simultaneously and on the same spatial and timescales. However, some processes will only affect certain phases of baryons in the galaxy. For example, the hydrodynamical effects (i.e., gas stripping, evaporation, etc.) will only affect the gas within the galaxy, while leaving the stars relatively unaffected. Conversely, gravitational effects (mergers, interactions, etc.) will affect both stars and gas, each of which should return to equilibrium over different timescales \citep{1962ApJ...136..748E,1984ApJ...276..127J}. We can exploit this difference in relaxation time to disentangle galaxy evolution processes that happen exclusively in clusters (hydrodynamical processes e.g., RPS, viscous stripping and evaporation) processes from those that would work in both groups and clusters (gravitational interactions e.g., mergers/interactions). Previous studies utilizing the stellar and ionised kinematics have focused on measuring rotational curves for galaxies in clusters and the offset of kinematic centres between the stellar and gas velocity fields \citep{1999AJ....118..236R,2008A&A...483..783K}, and idealized simulations have shown that the drag force on the stripped gas is sufficient to move the dynamical centre of both the stars and dark matter, with the stellar disk becoming thicker as a result of decreased gravitational potential \citep{2012MNRAS.420.1990S}.

In this work, we use stellar and ionised gas kinematic maps of field and cluster galaxies from the Sydney-AAO Multi Integral field (SAMI) Galaxy survey \citep{2015MNRAS.447.2857B,2017MNRAS.468.1824O} to investigate the effect a cluster environment has on the dynamics of galaxies. We use \textsc{kinemetry} \citep{2006MNRAS.366..787K} to model the dynamics of the stars and ionised gas, and measure the kinematic asymmetry. The structure of the paper is as follows: In Sect. \ref{data} we describe the SAMI Galaxy Survey, the data products used in this work, and our sample selection. In Sect. \ref{kinematics}, we describe how we model the stellar and ionised gas kinematics maps and how we quantify kinematic asymmetry. In Sect. \ref{results}, we introduce our results and discuss potential interpretations, before offering our conclusions in Sect. \ref{conc}. Throughout this work, we adopt a flat $\Lambda$CDM cosmology with $\Omega_M$=0.3, $\Omega_\Lambda$=0.7, with Hubble's constant $H_0$=70 kms$^{-1}$ Mpc$^{-1}$. 

\section{Data}

\begin{table*}
    \centering
    \begin{tabular}{c|c|c|c|c|c|c|c}
        \hline
        Cluster  & R.A. & Dec. & $z$ & $M_{200}$ & $R_{200}$ & $N_{\rm EGA}$ & $N_{\rm tot}$ \\
        \hline
        EDCC0442 & 6.380 & -33.046 & 0.0498 & 14.45 & 1.41 & 1 & 10 \\
        Abell0085 & 10.460 & -9.303 & 0.0549 & 15.19 & 2.42 & 6 & 31 \\
        Abell0119 & 14.067 & -1.255 & 0.0442 & 14.92 & 2.02 & 8 & 20 \\
        Abell0168 & 18.815 & 0.213 & 0.0449 & 14.28 & 1.33 & 5 & 18 \\
        Abell2399 & 329.389 & -7.794 & 0.0579 & 14.66 & 1.63 & 3 & 36 \\
        Abell3880 & 336.977 & -30.575 & 0.0578 & 14.64 & 1.62 & 1 & 20 \\
        APMCC0917 & 355.397 & -29.236 & 0.0509 & 14.26 & 1.19 & 0 & 6 \\
        Abell4038 & 356.938 & -28.140 & 0.0293 & 14.36 & 1.46 & 2 & 13 \\
        Total & - & -  & -  & -  & - & 26 & 154 \\
        \hline
        GAMA & - & - & - & - & - & 81 & 751 \\
        \hline
    \end{tabular}
    \caption{Properties for clusters from which we selected our sample, as well as the number of galaxies selected from each cluster and the GAMA sample. Values are taken from \citealt{2017MNRAS.468.1824O}. Col. 1: Cluster Name, Col. 2: Right Ascension (J2000), Col. 3: Declination (J2000), Col. 4: Redshift, Col. 5: Virial mass [$\log(M_{200}/M_\odot)$], Col. 6 Virial radius [Mpc], Col. 7 Number of Elevated Gas Asymmetries (EGA) galaxies in each cluster, Col. 8 Total number of galaxies selected from each cluster, or sample.}
\end{table*}

\label{data}
\subsection{The SAMI Galaxy Survey}
The SAMI Galaxy Survey is an integral field spectroscopic survey of galaxies within the local Universe, probing galaxies over a large range of stellar mass, morphology and galactic environments \citep{2015MNRAS.447.2857B,2016MNRAS.463..170C,2017MNRAS.468.1824O,2017ApJ...844...59B}. Spectral cubes were obtained using the SAMI instrument \citep{2012MNRAS.421..872C} on the 3.9m Anglo-Australian Telescope connected to the AAOmega spectrograph \citep{2006SPIE.6269E..0GS}. The SAMI instrument had 13 hexabundles and 26 sky fibres over a 1-degree field-of-view. Each hexabundle consisted of 61 optical fibers with a diameter of 1.6 arcsec, giving each hexabundle a diameter of 15 arcsec. The SAMI Galaxy Survey targeted galaxies within the Galaxy and Mass Assembly Survey \citep[GAMA;][]{2011MNRAS.413..971D} fields G09, G12 and G15, and eight rich galaxy clusters with virial masses within the range 14.25<$\log(M_{200}/M_\odot)$<15.19. For this work, we utilize both the field and cluster galaxies. For each cluster, member galaxies are selected following the method described in sect. 4 of \citep{2017MNRAS.468.1824O}. Data products used in this work (stellar and gas velocity maps, emission line fluxes, etc.), are described in detail in the SAMI Data Releases \citep{2018MNRAS.475..716G,2018MNRAS.481.2299S,2021MNRAS.505..991C} and the Cluster Redshift Survey paper \citep{2017MNRAS.468.1824O}. Those are summarized below for completeness.

Briefly, stellar kinematics are extracted using a penalized pixel fitting algorithim \citep[\textsc{pPXF};][]{2004PASP..116..138C} which models the stellar continuum as a linear superposition of stellar templates. All stellar kinematics are obtained using additive Legendre polynomials of degree $n=12$. Regions of nebular and strong sky-line emission are masked. A first round of fitting is performed on a set of elliptical annuli to derive a subset of optimal stellar templates that are subsequently used to fit individual spaxels. Starting from the centre, the procedure creates 5 equally spaced elliptical annuli using spaxels with a SNR$\geq$3. Where annuli do not meet the nominal SNR requirement of 25 \AA$^{-1}$, annuli are combined outside-in until this threshold is met. The procedure then fits individual spaxels using the corresponding restricted subset of templates used to model the annular spectra \citep{2017ApJ...835..104V}. Ionised gas velocity maps are constructed using \textsc{lzifu} \citep{2016Ap&SS.361..280H} after the stellar continuum was modelled and subtracted as described in \cite{2019ApJ...873...52O}. Emission lines are modelled as 1, 2 or 3 component Gaussians to capture different kinematic features \citep{2014MNRAS.444.3894H,2016Ap&SS.361..280H} for each continuum-subtracted spaxel's spectrum. This work uses the 1-component velocity maps throughout. 

Effective radii $R_{\rm e}$ for field, group and cluster galaxies are described in \cite{2021MNRAS.504.5098D,2024MNRAS.532.1775D}. Briefly, $R_{\rm e}$ are measured using multi-Gaussian expansion \citep[MGE;][]{1994A&A...285..723E} photometric modeling on images from Sloan Digital Sky Survey (SDSS) Data Release 7 \citep{2009ApJS..182..543A}, reprocessed as described in \cite{2011MNRAS.412..765H}. Cluster galaxy $R_{\rm e}$ are similarly measured using images from SDSS Data Release 9 \citep{2012ApJS..203...21A} and VLT Survey Telescope ATLAS Survey \cite[VST;][]{2013Msngr.154...38S,2015MNRAS.451.4238S}. 

Following \cite{2013MNRAS.435.2903B}, 5$^{\rm th}$ nearest neighbour surface density values for both field and cluster galaxies are calculated using $\Sigma_5=5/\pi d_5^2$ where $d_5$ is the distance to the 5$^{\rm th}$ nearest neighbour within $\pm$1000 kms$^{-1}$ of the SAMI target redshift. For field galaxies, redshifts are taken from GAMA, whereas cluster galaxies uses redshifts form the SAMI Cluster Redshift Survey. Galaxies included in the density estimates have absolute $r-$band magnitudes of $M_r<-18.6$ or $M_r<-19$, with the $M_r<-19$ requirement being included to compute densities for secondary targets at $z>0.1$ \citep{2017ApJ...844...59B}.

Concentration values $C(\rm H\alpha)$ are derived in a similar manner to \cite{2017MNRAS.464..121S} and are described in \cite{2019ApJ...873...52O}. Briefly, $\rm H\alpha$ and $r$-band cumulative flux profiles using elliptical isophotes, where emission lines within the $r$-band are masked, are measured to determine the radius containing 50\% of the $H\alpha$ and $r$-band flux within the SAMI bundle. The concentration is determined by $C(H\alpha)$=$r_{50,\rm H\alpha}/r_{50,r}$. Stellar masses are computed using rest-frame $g-i$ colours and $i-$band absolute magnitudes following the method in \cite{2011MNRAS.418.1587T}, assuming a \cite{2003PASP..115..763C} Initial Mass Function and exponentially declining star-formation history.  

\subsection{Sample Selection}
\begin{figure*}
    \centering
    \includegraphics[width=1\linewidth]{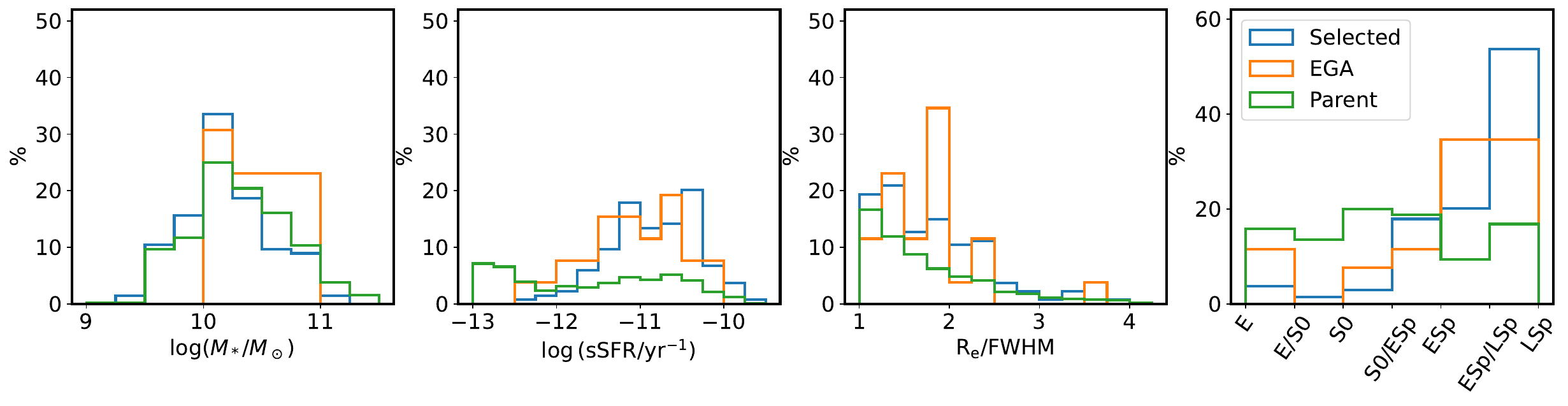}
    \caption{Histograms of stellar mass (left), specific star formation rate (sSFR, middle-left), $R_{\rm e}$/FWHM (middle-right) and morphology from visual classification (right) for the cluster galaxies in our sample (blue line) and the subsample of those galaxies with elevated gas asymmetries (EGA galaxies; orange line) and the parent SAMI Cluster sample (green line).}
    \label{histogram}
\end{figure*}

We select galaxies from the GAMA regions as well as clusters. We exclude galaxies where maximum stellar continuum SNR/spaxel<3 and SNR(H$\alpha$)<20 within 2$R_{\rm e}$, or within the FOV of the SAMI IFU for galaxies not sampled to $2R_{\rm e}$. Galaxies that meet the stellar continuum criterion, but not the H$\alpha$ criterion are included as ionised gas non-detections. Similarly, galaxies that meet the H$\alpha$ criterion, but not the stellar continuum criterion are included as stellar non-detections. We also exclude galaxies where $R_{\rm e}$ is less than the estimated seeing (i.e., $R_{\rm e}$/FWHM>1). Finally, we use the [S~{\sc ii}]-BPT diagnostic to remove non-star forming galaxies using 1$R_{\rm e}$ integrated fluxes, and the criteria stipulated in \citep{2006MNRAS.372..961K}. The [S~{\sc ii}]-BPT was used as it allows a cleaner separation between star-forming, AGN and LINER galaxies, than the [N~{\sc ii}]-BPT alone \citep{2001ApJ...556..121K,2016MNRAS.461.3111B,2021ApJ...915...35L}. This removes galaxies where the emission is mostly due to non-SF sources (e.g., AGN, shocks), which can cause the LOSVD to not be adequately modeled with a single component Gaussian \citep{2013MNRAS.436.2576L,2015ApJ...806...84L,2017ApJ...834...30F}. While SAMI does have ionised gas velocity maps with multiple components, the high SNR criteria necessary to model multiple components removes many spaxels that are necessary to model 2D kinematic maps (See Sect. 4.1 of \citealt{2018MNRAS.475..716G}). Because of this, we opt to use the single component Gaussian maps; we expand on this in the next section. Once a galaxy satisfies these criteria, spaxels within 2$R_{\rm e}$ in the stellar $v_{\rm los}$ and $\sigma$ maps with stellar continuum SNR<3 are masked. A similar requirement is applied to the ionised gas $v_{\rm los}$ and $\sigma$ maps using SNR(H$\alpha$). Our sample selection is visually represented using histograms of stellar mass, specific star formation rate (sSFR), $R_{\rm e}$/FWHM and visual classification in Fig. \ref{histogram}. Our selection criteria primarily selects star-forming, massive ($\log(M_*/M_\odot$)>10) disk galaxies from the parent cluster sample. We are necessarily selecting against low sSFR systems due to our H$\alpha$ SNR requirement, meaning we are biasing against quenched systems. Our criteria reduces the galaxies from cluster and GAMA regions to samples of 195 and 883 galaxies, respectively.

\subsection{Kinematic Analysis}

\begin{figure}
    \centering
    \includegraphics[width=1\linewidth]{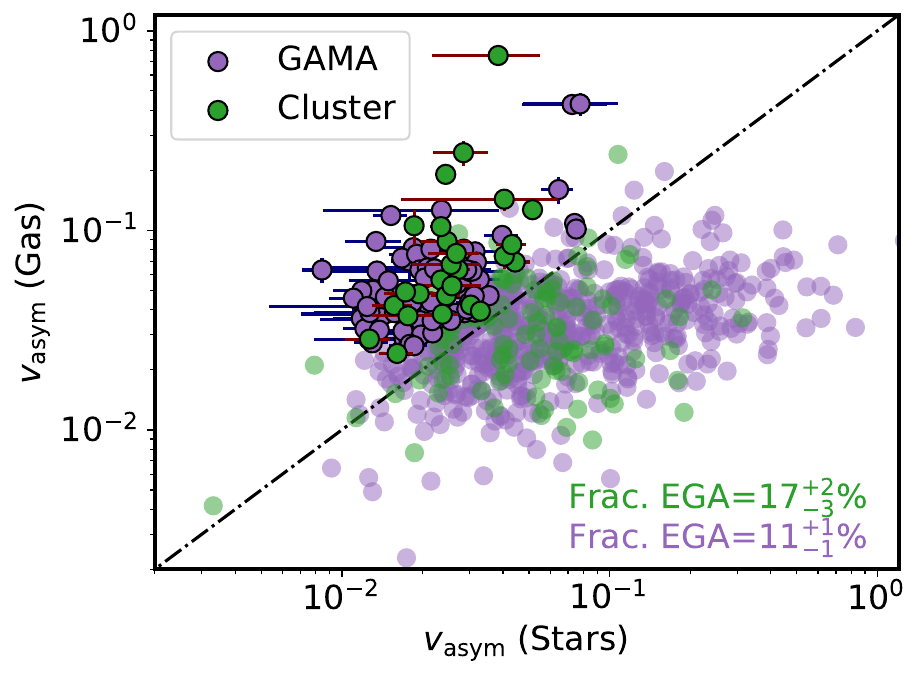}
    \caption{$v_{\rm asym}$ (Stars) vs. $v_{\rm asym}$ (Gas) for GAMA (purple) and cluster galaxies (green). Elevated gas asymmetry (EGA) galaxies in each samples are have solid colours, while non-EGA galaxies are faded. The dot-dashed line is a 1:1 line. We find that the fraction of EGA galaxies in cluster environments is 17$^{+2}_{-3}$\%, whereas the fraction of EGA galaxies in field environments is 11$^{+1}_{-1}$\%.}
    \label{gas_stars_marginal}
\end{figure}

\begin{figure}
    \centering
    \includegraphics[width=\linewidth]{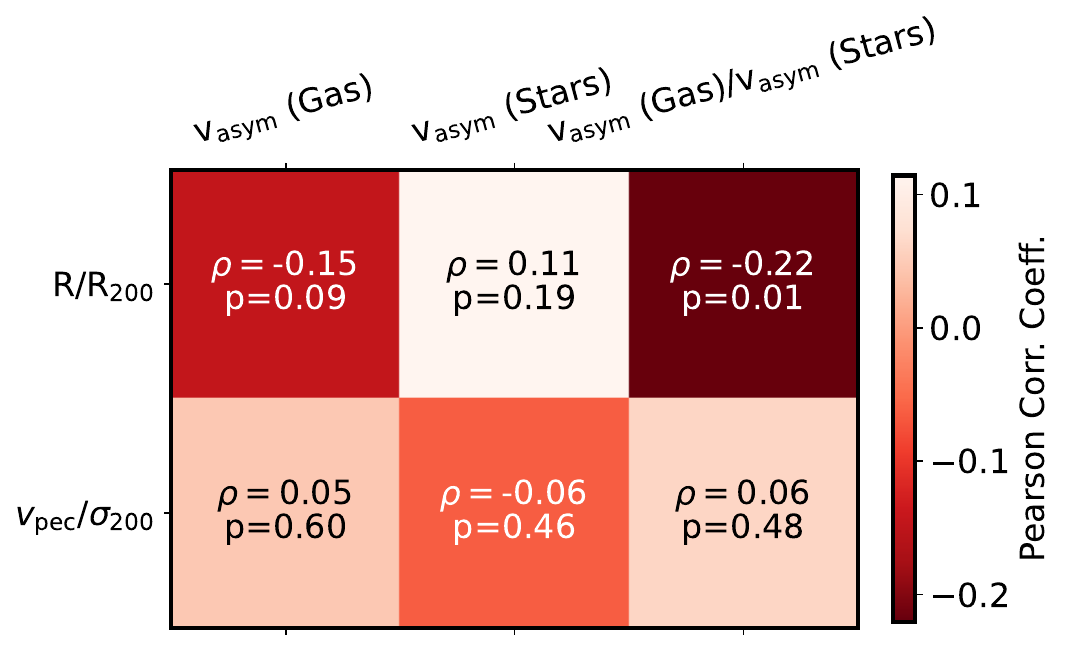}
    \caption{Grid plot showing Pearson correlation coefficients between $v_{\rm asym}$(Gas), $v_{\rm asym}$(Stars), R/R$_{200)}$, $v_{\rm pec}/\sigma_{200}$. There is no particularly strong trend or significant trends with $v_{\rm asym}$ and $v_{\rm pec}/\sigma_{200}$. However, there is a weak correlation between R/R$_{200}$ and $v_{\rm asym}$(Gas)/$v_{\rm asym}$(Stars).}
    \label{corr_grid}
\end{figure}

\begin{figure*}
    \centering
    \includegraphics[width=1\linewidth]{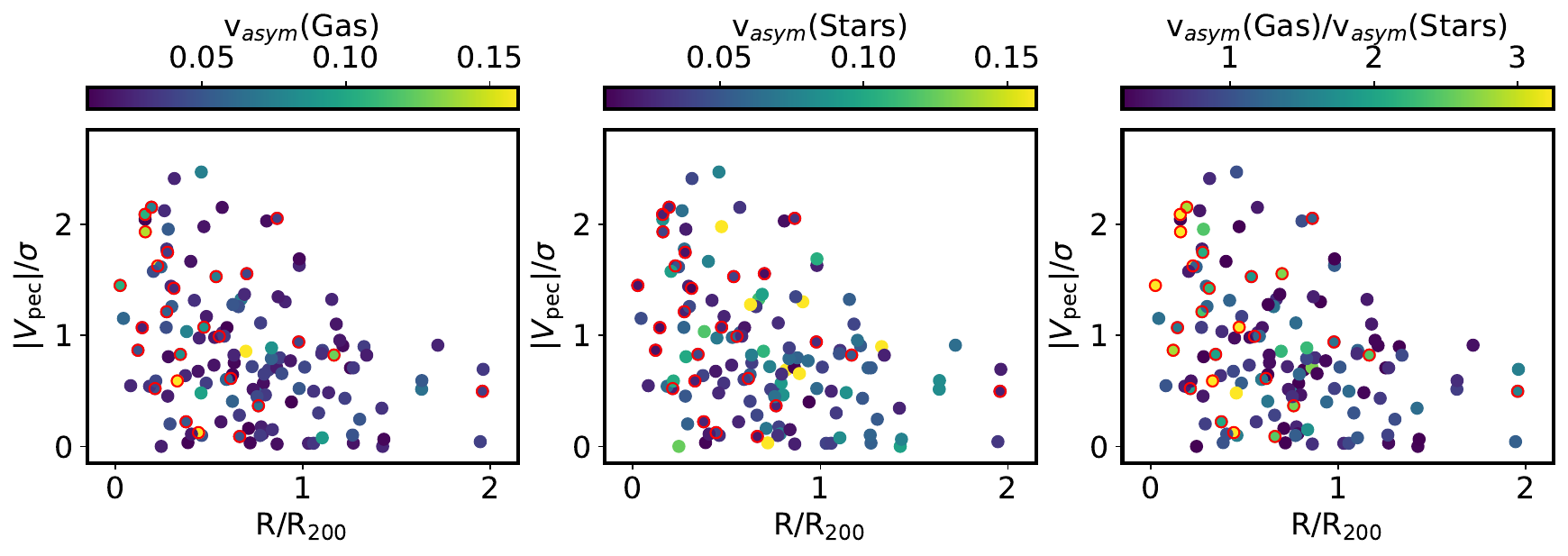}
    \caption{Projected phase space (PPS) of SAMI cluster galaxies within our sample. Symbols are colour-coded according to their $v_{\rm asym}$ (Gas), $v_{\rm asym}$ (Stars) and the ratio of $v_{\rm asym}$ (Gas) and $v_{\rm asym}$ (Stars) in the left, centre and right panel, respectively. Symbols for elevated gas asymmetry (EGA) galaxies are marked with red edges.}
    \label{PPS_gas_over_stars}
\end{figure*}

\label{kinematics}
\textsc{kinemetry} is an algorithm that extends photometric analysis used for surface brightness photometry to line-of-sight kinematic maps $v_{\rm los}$ and line-of-sight velocity dispersion $\sigma$ maps. \textsc{kinemetry} constructs models of kinematic maps by fitting a series of concentric ellipses with position angle (PA) and axial ratio ($q=b/a$), where $b$ and $a$ are the semi-minor and semi-major axes. Like other tilted-ring fitting algorithms \citep{1974ApJ...193..309R,1987hrcs.book.....B,2007ApJ...664..204S,2015MNRAS.451.3021D}, \textsc{kinemetry} fits a Fourier Series to the ellipse,
\begin{equation}
    K(a,\theta) = A_0 + \sum_{m=1}^{m=N} A_m\sin{(m\theta)} + B_m\cos{(m\theta)},
    \label{FS}
\end{equation}
where $a$ is the semi-major axis of the ellipse, $\theta$ is the azimuth along the ellipse with respect to the semi-major axis, A$_0$ is the zeroth harmonic term and A$_m$ and B$_m$ are the $m\rm{th}$ additional harmonic terms.

Equation \ref{FS} can be represented more compactly as:
\begin{equation}
    K(a,\theta) = A_0 + \sum_{m=1}^{m=N} k_m\cos{(m[\theta - \phi_m(a)])},
    \label{compactFS}
\end{equation}
where $k_m = \sqrt{A_m^2+B_m^2}$ and $\phi_n = \arctan\frac{A_m}{B_m}$. The $k_m$ parameters describe the kinematics (i.e., $V_{\rm rot}= k_1$) of the galaxy while PA and $q$ describe the geometry of the ellipse. A rotating, thin disk galaxy can be described with a single cosine term (i.e., $V(r,\theta) = V_{\rm rot}(r)\cos{\theta}$), with deviations from circular motions resulting in an asymmetric function along sampled ellipses. These `asymmetries' will be encoded in the $m>1$ terms in Equation \ref{compactFS}. 

Following \citet{2024MNRAS.531.3011B}, we normalise the higher-order terms to $S_{05}=\sqrt{0.5V_{\rm rot}^2+\sigma^2}$. $V_{\rm rot}$ is taken where the rotation curve reaches its maximum, and $\sigma$ is the light-weighted average within 1$R_{\rm e}$. $V_{\rm rot}$ is corrected for inclination by $v_{\rm los}$/$\sin{i}$, where $i$ is the inclination of the galaxy. Mathematically, $v_{\rm asym}$ is given by: 
\begin{equation}
\begin{aligned}\label{eq:vasym}
v_{\rm asym} = \frac{k_2+k_3+k_4+k_5}{4S_{05}}.
\end{aligned}
\end{equation}
As demonstrated in \citet{2024MNRAS.531.3011B}, this definition is valid for both rotation and pressure supported galaxies. To measure $V_{\rm rot}$ and $\sigma$, we use \textsc{kinemetry} on the respective maps by fixing the PA and $q$ to the global kinematic PA (PA$_{\rm kin}$) and $q$. PA$_{\rm kin}$ is measured using the \textsc{FitKinematicPA} routine, where spaxels with SNR<3 are ignored. Finally, the inclination of the galaxy is determined using,
\begin{equation}
\begin{aligned}
\label{inc_corr}
\cos^2{i} = \frac{q^2-q_0^2}{1-q_0^2},
\end{aligned}
\end{equation}
where $q_0$ is the intrinsic (3D) axis ratio. The chosen $q_0$ is dependent on the morphology of the galaxy. We select $q_0=0.6$ and $q_0=0.2$ for galaxies with E/S0, and Sa/Sb/Sc morphologies, respectively. These values have previously been used for inclination corrections in SAMI galaxies \citep[e.g.,][]{2016MNRAS.463..170C}, and also are consistent with those used in \cite{2024MNRAS.527.7438R}, who used a combination of T-type values (ranging from -3 to 10) and morphologies determined from deep learning models trained on MaNGA DR17 images \citep{2019MNRAS.483.2057F,2021AAS...23811901D} along with a prescription in \citep{1983A&A...118....4B} to determine $q_0$. Their values for $q_0$ ranged from 0.15 and 0.55 for disk and spheroidial galaxies, respectively.

To evaluate kinematic asymmetries, we run \textsc{kinemetry} out to 1$R_{\rm e}$ in radial steps of half the measured seeing. \textsc{kinemetry} was originally used on kinematic maps with high SNR and spatial resolution. In contrast, galaxies in our sample do not necessarily have these qualities, which can lead to geometric parameters varying chaotically during the fitting. To prevent this, when measuring kinematic asymmetries we fix the PA to the PA$_{\rm kin}$ and $q$ to the MGE derived $q$ value. Uncertainties on the asymmetries are estimated using a 100 Monte Carlo realizations, where we re-run \textsc{kinemetry} on the same velocity map with Gaussian noise injected into each spaxel corresponding to the error on the velocity measurements. Final uncertainties correspond to the standard deviation of the Monte Carlo distribution. Occasionally, our spaxel SNR requirements create discontinuities along the ellipse that prevents \textsc{kinemetry} from computing the asymmetries. To circumvent this, we follow the method in App. A of \cite{2023PASA...40...60B}, where we replace the masked spaxel with the median value of the adjacent 8 spaxels. We remove 41 and 132 galaxies from the cluster and GAMA samples, respectively, where we replaced more than 70\% of the spaxels along the sampled ellipse. The mean value of replaced spaxels, in both the ionsied gas and stellar kinematic maps, for our sample after our selection criteria is 3\%. This brings our final sample to 154 cluster galaxies and 751 GAMA galaxies.

\section{Results \& Discussion}
\label{results}
In this section, we present our results and discuss how passage through the cluster affects the kinematics of the ionised gas and stars in galaxies, by examining where galaxies with large kinematic asymmetries are in projected phase space (PPS) as well as their star-formation activity and morphology.

To investigate galaxies \textit{primarily} experiencing hydrodynamical cluster effects, we focus most of our analysis on galaxies with elevated gas asymmetries compared to stellar asymmetries, henceforth referred to as EGA galaxies. Hydrodynamical cluster effects should predominantly impact the gas dynamics, and may thus be traced through elevated gas asymmetries. We expect the comparison between $v_{\rm asym}$(Gas) and $v_{\rm asym}$(Stars) to be fair since the $S_{05}$ for both the stars and gas are similar (e.g., $S_{05,\rm Gas}/S_{05,\rm Stars}\approx 1$). We select galaxies as EGA galaxies where [$v_{\rm asym}$(Gas) - $v_{\rm asym}$(Stars)] >$3\times\sqrt{Err[v_{\rm asym}(Gas)]^2 + Err[v_{\rm asym}(Stars)]^2)}$, where $Err[v_{\rm asym}]$ is the Monte Carlo estimated errors for each asymmetry measure. EGA galaxies represent 26/154, or 17$^{+2}_{-3}$\% of the cluster sample and 72/751, or 11$^{+1}_{-1}$\% of the GAMA sample. Fig. \ref{gas_stars_marginal} show the $v_{\rm asym}$(Gas)-$v_{\rm asym}$(Stars) space for the cluster and GAMA samples. Comparing the fractions of EGA galaxies in the cluster and GAMA samples, we find that the fraction of cluster galaxies with elevated gas asymmetries is more than 1$\sigma$ higher than in GAMA. Flux and velocity maps for each galaxy in our sample are in App. \ref{flux_velo_images}.

\subsection{Projected Phase Space \& Environment}
\label{PPS}
Projected Phase Space (PPS; $v_{\rm pec}$/$\sigma_{200}$ vs. $R/R_{200}$) is a useful tool to examine where galaxies are within the cluster environment, and how this drives their evolution \citep[e.g.,][]{1982AJ.....87..945K,2005MNRAS.356.1327G,2013ApJ...768..118N,2015MNRAS.448.1715J,2016MNRAS.463.3083O,2017ApJ...843..128R, 2019ApJ...873...52O,2019MNRAS.484.1702P,2021MNRAS.503.3065S,2024A&A...691A.135B}. As galaxies begin to fall into the cluster, the radial velocity will increase, peaking as it approaches the pericentre of its orbit. Once it passes pericentre, the velocity will decrease as the galaxy approaches the outskirts of the cluster, where the galaxy will spend the majority of its orbit \citep{2005MNRAS.356.1327G,2013MNRAS.431.2307O}.

Correlation coefficients and $p$-values between the asymmetry, $v_{\rm pec}/\sigma_{200}$ and R/R$_{200}$ are shown in Fig. \ref{corr_grid} and Fig. \ref{PPS_gas_over_stars} shows the PPS for our sample of SAMI cluster galaxies, with points coloured by $v_{\rm asym}$ (Stars) (left), $v_{\rm asym}$ (Gas) (middle) and $v_{\rm asym}$ (Gas)/$v_{\rm asym}$ (Stars) (right). The left and middle panels of Fig. \ref{PPS_gas_over_stars}, do not show any particularly strong trend with either $v_{\rm asym}$(Gas) or $v_{\rm asym}$(Stars) and $R/R_{200}$ or $v_{\rm los}$/$\sigma_{200}$, which the correlation cofficent in Fig. \ref{corr_grid} confirms. There is a weak, but significant, correlation ($\rho$=-0.22,p=0.01) between $v_{\rm asym}$(Gas)/$v_{\rm asym}$(Stars) and $R/R_{200}$.

\begin{figure}
    \centering
    \includegraphics[width=1\linewidth]{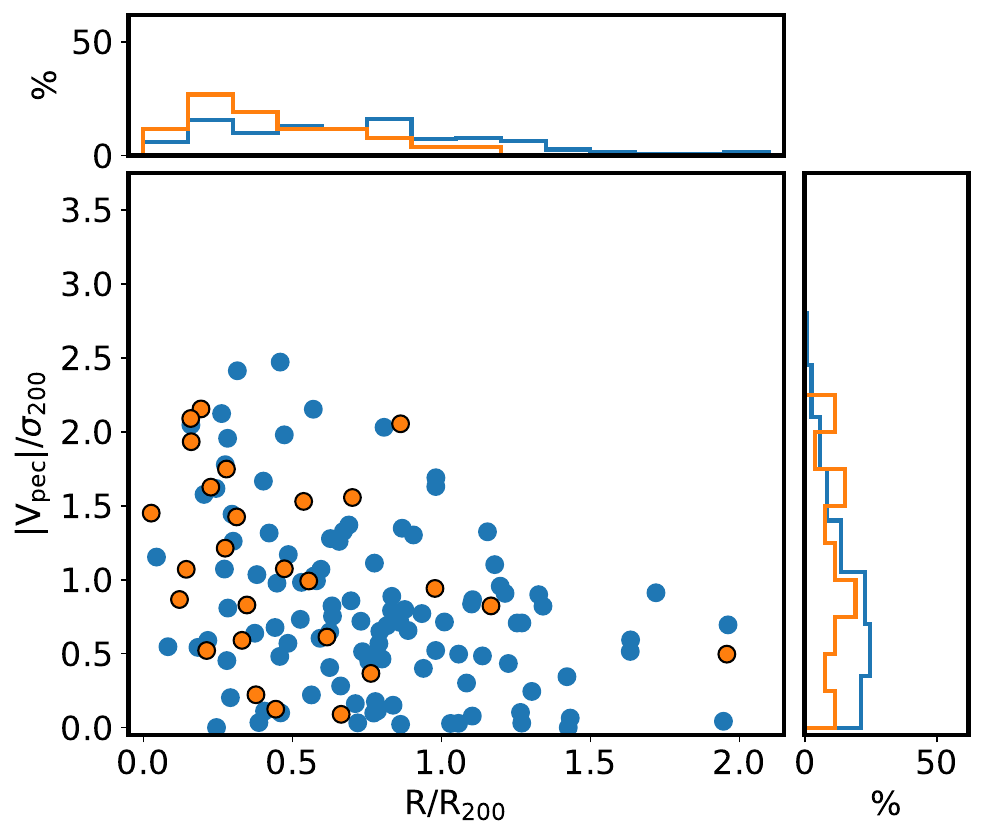}
    \caption{PPS diagram for galaxies within our sample (blue) with galaxies with elevated gas asymmetries (EGA galaxies) marked in orange. Marginal distributions for $R/R_{200}$ and $v_{\rm pec}/\sigma$ are shown above the x-axis and next to the y-axis, respectively. EGA galaxies in clusters are almost exclusively found within the cluster, barring 2 galaxies}
    \label{PPS_marginal}
\end{figure}

EGA galaxies in our sample are almost exclusively found within the cluster virial radius ($R/R_{200}$<1, see top panel of Fig. \ref{PPS_marginal}), barring 2 galaxy outside $1R/R_{200}$. Within $1R/R_{200}$, there are even number of galaxies (13/26) with high-$v_{\rm pec}$ velocities ($v_{\rm pec}/\sigma$>1) with low-$v_{\rm pec}$ ($v_{\rm pec}/\sigma$<1). 

\begin{figure*}
    \centering
    \includegraphics[width=1\linewidth]{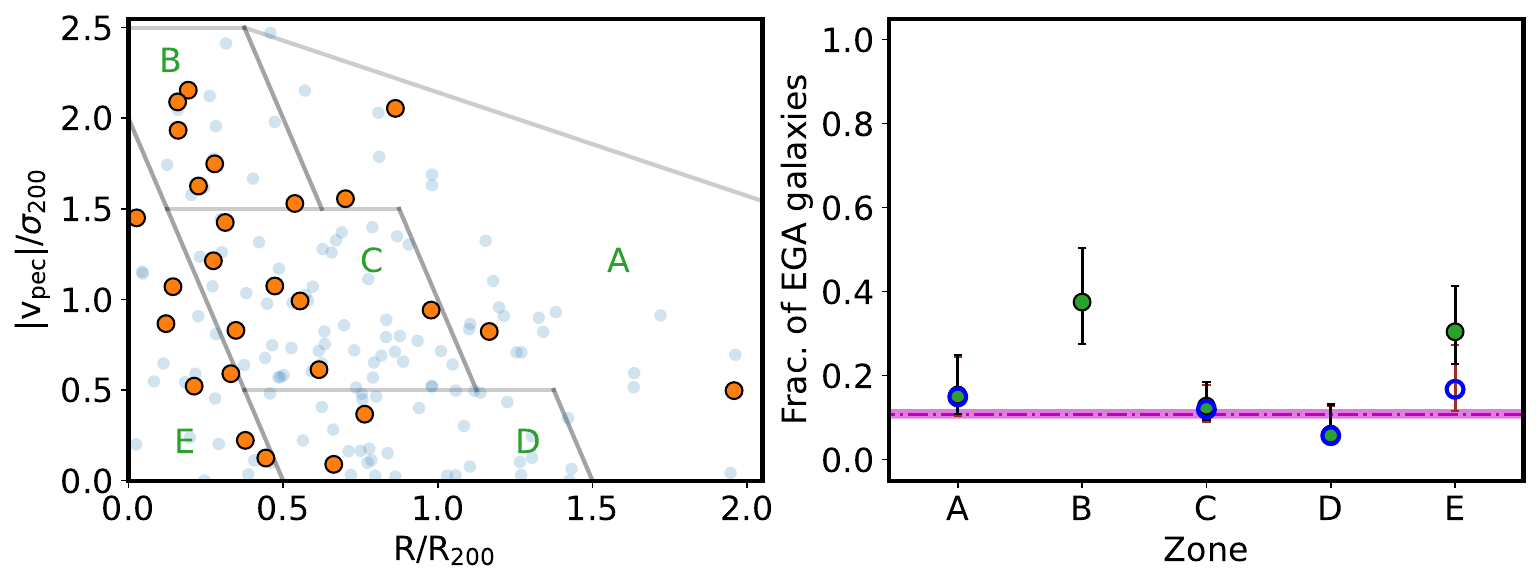}
    \caption{\textit{Left}: Projected Phase Space (PPS), but with discrete regions defined in \citep{2017ApJ...843..128R} corresponding to infall time. Zone A corresponds to the first infalling galaxies, Zone B corresponds to recent infalling population, Zone C corresponds intermediate regions, Zone D corresponds to ancient/intermediate regions, and Zone E corresponds to ancient infallers. \textit{Right}: The fraction of EGA galaxies within each of the zones listed in \textit{left}. The open blue circles are the fraction after removing possible contamination from other zones. The magenta line represents the fraction of EGA galaxies expected when not in cluster. When accounting for the contamination of recent infallers in Zone E, we find the fraction of EGA galaxies is consistent with fraction of EGA galaxies in non-cluster environments.}
    \label{PPS_infall}
\end{figure*}

We use the discrete regions of PPS from \cite{2017ApJ...843..128R} to estimate the stage of infall for the EGA galaxies. The authors defined the regions to maximize the fraction of galaxies with a similar infall time, with recent, recent-intermediate, intermediate and ancient infallers having mean infall times $T_{\rm in}$ of 1.61, 1.97, 4.98 and 8.35 Gyrs, respectively. We find that the high-$v_{\rm pec}$ EGA galaxies are in the recent infaller region (Zone B in Fig.\ref{PPS_infall}), while the low $v_{\rm pec}$ are predominately located within the intermediate infaller region (Zone C in Fig.\ref{PPS_infall}). The location and estimated $T_{\rm in}$ would suggest the high-$v_{\rm pec}$ EGA galaxies are experiencing their first pericenter passage as they move through the cluster. Based on their proximity to $R_{200}$, the low-$v_{\rm pec}$ EGA galaxies are probably approaching the outskirts of the cluster following their first pericenter passage (i.e., backsplash galaxies). The right panel of Fig. \ref{PPS_infall} shows the fraction of EGA galaxies located within each zone, we find that the fraction of EGA galaxies in Zone B and Zone E are 2$\sigma$ above the fraction within the GAMA sample ($38^{+10}_{-12}$\% and $30^{+7}_{-10}$\% vs. $11^{+1}_{-1}$\%). We also find that the fractions of EGA galaxies in Zones A, C and D are consistent with fraction of EGA galaxies expected when galaxies are not within a cluster. The higher fractions in Zone B \& E suggests that there is something specific to the cluster environment that is causing the galaxies within these Zones to be EGA galaxies.

While useful, PSS can be difficult to interpret due to it projecting three dimensions in 2D. For example, in 3D, a galaxy could have a large radial velocity and small $R/R_{200}$ (so infalling into the cluster), but projected at the line-of-sight, giving it a low $v_{\rm pec}/\sigma_{200}$ and making it appear as an ancient infaller, when it is actually completing its first pericentre passage. This contamination becomes increasingly important at small $R/R_{200}$ and we want to delineate between ancient and recent infallers. To understand how significant this contamination can be, \cite{2021MNRAS.500.1784D} trained different machine learning model trained on the 3D orbital histories of simulated galaxy clusters. With their best performing model, the authors found that as much as 45\% of ancient infallers (Zone E) can be misidentified recent infallers (Zone B) in PPS. Adopting their estimates for contamination of recent infallers within each zone from fig. 6 in \citep{2021MNRAS.500.1784D}, we correct the fraction of EGA galaxies in each zone. (i.e., Frac(EGA,corr) = Frac(EGA) $\times 1-Pr(\rm RIN)$, where $Pr(\rm RIN)$ is the probability that galaxies in each zone is a recent infaller). The probabilities are quoted in Tab. \ref{Probs}. Once removing contamination from each zone, we find that the fraction of EGA galaxies in zone A, C, D are consistent with the fraction of EGA galaxies found in non-cluster environments, and there is no longer a significant difference in Zone E. This suggests that galaxies being EGA is due to their location in Zone B.

\begin{table}
    \centering
    \begin{tabular}{c|c|c|c|c|c}
         \hline
         \hline
         & Zone A & Zone B & Zone C & Zone D & Zone E  \\
         \hline
        Pr$(\rm RIN)$ & 0.03  & - & 0.06 & 0.06 & 0.45 \\
         \hline
         \hline
    \end{tabular}
    \caption{Probabilities that galaxies in each zone are actually recent infallers (i.e., belong in Zone B) from \citealt{2021MNRAS.500.1784D}.}
    \label{Probs}
\end{table}

\begin{figure*}
    \centering
    \includegraphics[width=\linewidth]{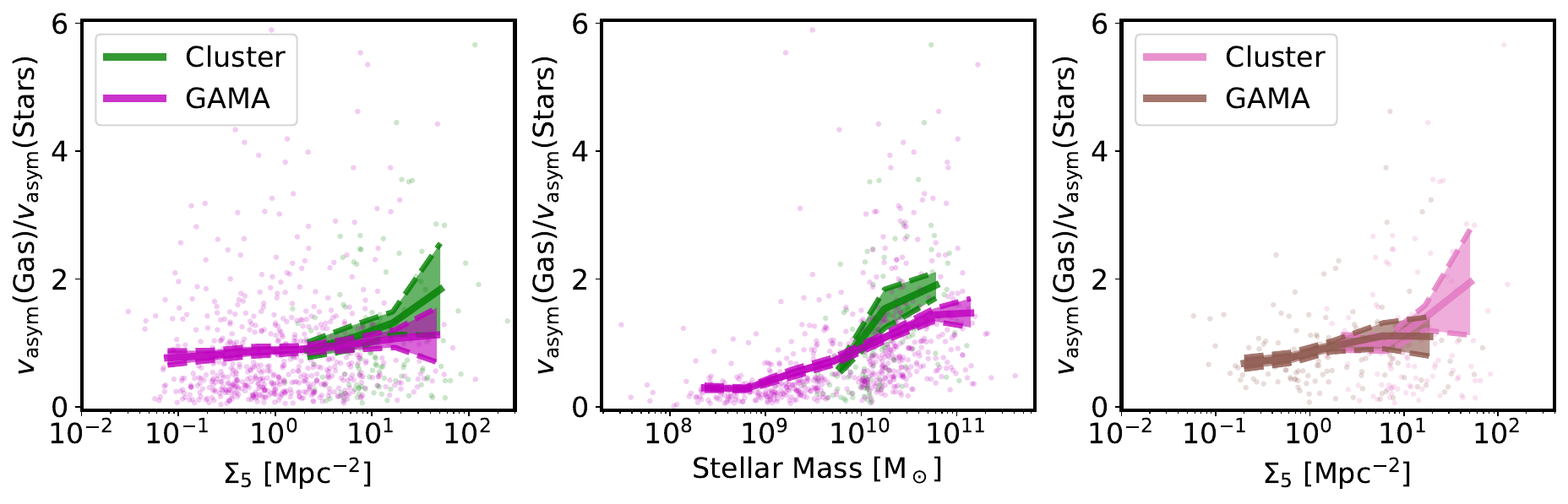}
    \caption{\textit{Left:} $v_{\rm asym}$(Gas)/$v_{\rm asym}$(Stars) vs. $\Sigma_5$ for cluster galaxies (green) and GAMA galaxies (purple). \textit{Centre}: $v_{\rm asym}$(Gas)/$v_{\rm asym}$(Stars) vs. stellar mass for cluster galaxies (green) and GAMA galaxies (purple) \textit{Right}: The same as \textit{left} after matching stellar mass and redshift distributions, with cluster galaxies in pink and GAMA galaxies in brown. The solid line represents the average $v_{\rm asym}$(Gas)/$v_{\rm asym}$(Stars) in bins of $\Sigma_5$, or stellar mass 0.5 dex wide. Bins are only plotted if there are 5 or more galaxies in each bin. At $\Sigma_5$>10 Mpc$^{-2}$, galaxies in cluster have larger $v_{\rm asym}$(Gas)/$v_{\rm asym}$(Stars); however, this difference becomes less significant after matching stellar mass and redshift distributions.}
    \label{sigma5_mass_match}
\end{figure*}

To isolate the impact of the cluster environment itself on gas asymmetries, we compare $v_{\rm asym}$ (Gas)/$v_{\rm asym}$(Stars) against $\Sigma_5$ in cluster and field galaxies (Fig. \ref{sigma5_mass_match}). This comparison will help delineate environmental effects exclusive to clusters (RPS, evaporation etc.) from those that happen in similarly dense non-cluster environments (e.g., interacting groups members in close proximity to each other). Mass and environment are correlated quantities, with more massive galaxies being found in denser environments, meaning our environmental trends could be driven by stellar mass rather than specifically environment \citep{2010ApJ...721..193P,2017ApJ...844...59B,2021MNRAS.508.2307V}. To mitigate this, we match the stellar mass and redshift distributions between the GAMA and cluster samples. We match the distributions in stellar mass bins ranging from $\log(M_*/M_\odot)=[9.5,12]$ with bins sizes of 0.5 dex, and redshift bins ranging from $z=[0,0.1]$ in bins of 0.01. Fig. \ref{sigma5_mass_match} shows $v_{\rm asym}$ (Gas)/$v_{\rm asym}$(Stars) against $\Sigma_5$ for the matched samples. Before matching in stellar mass and redshift, we find that at $\Sigma_5$>10 Mpc$^{-2}$, galaxies within a cluster have larger $v_{\rm asym}$ (Gas)/$v_{\rm asym}$ (Stars) than field galaxies, although this is not a significant difference. We similarly find that cluster galaxies will have a larger $v_{\rm asym}$(Gas)/$v_{\rm asym}$(Stars) at a given stellar mass. After matching in stellar mass and redshift, the difference between GAMA and cluster galaxies becomes even less obvious (see right panel of Fig. \ref{sigma5_mass_match}).

The EGA galaxies within Zone B are likely experiencing RPS. The RPS experienced by a galaxy in a cluster will be greatest when it reaches pericentre (i.e., large $v_{\rm pec}/\sigma_{200}$, small $R/R_{200}$), and this likely leads to disturbed kinematics in the gas, which can explain why a significant number of galaxies in Zone B are EGA galaxies. It is likely that the elevated gas asymmetries are primarily driven by RPS rather than gravitational interactions since the $v_{\rm asym}$ (Stars) value is low (See middle panel of Fig. \ref{PPS_gas_over_stars}), and gravitational interactions would cause elevated stellar asymmetries along with elevated gas asymmetries. As these galaxies leave the pericentre and their velocities slow down, the RPS will decrease ($F_{\rm RPS}\propto v_{\rm pec}^2$). This could explain why the fraction of EGA galaxies in other zones is consistent with the fraction of EGA galaxies found in GAMA, and suggests that EGA galaxies are solely due to RPS. However, it is likely that these EGA galaxies will continue to experience some gas stripping separate from RPS (e.g. viscous, evaporation, etc) while within the cluster, which could also explain the population of EGA galaxies outside Zone B. \cite{2007MNRAS.380.1399R, 2008MNRAS.388..465R} conducted hydrodynamical simulations of a galaxy within a cluster and found that RPS alone was not enough to strip galaxies down to expected radii, and that viscous stripping is needed to continually strip the gas within the galaxy.

\subsection{Star-Formation Activity}
\label{SFA}
\begin{figure}
    \centering
    \includegraphics[width=1\linewidth]{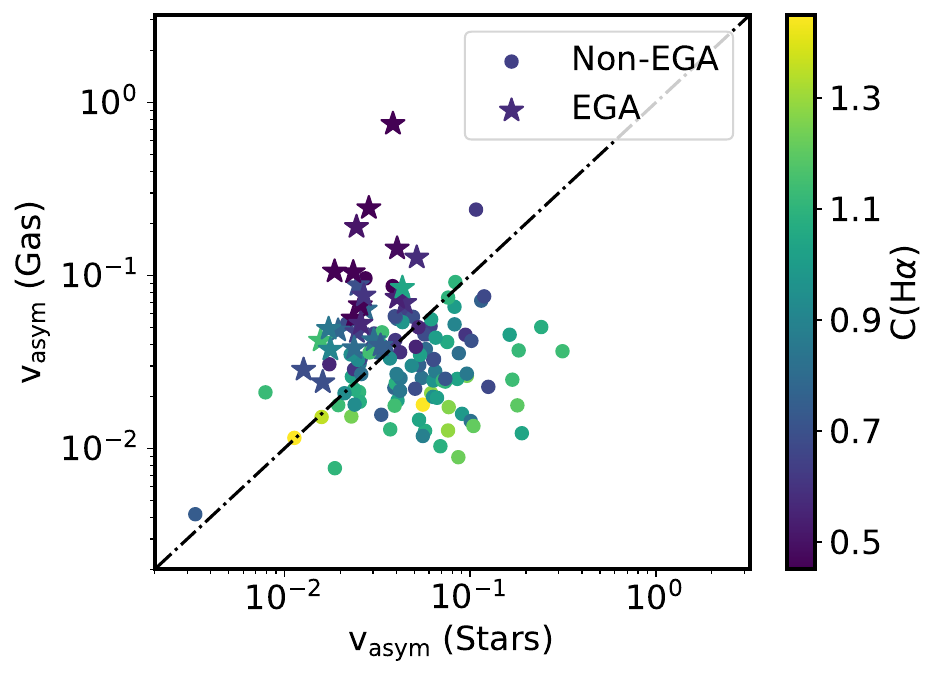}
    \caption{$v_{\rm asym}$ (Stars) vs. $v_{\rm asym}$ (Gas) coloured by $R_{50,\rm{H\alpha}}/R_{50,\rm{cont}}$. EGA galaxies are marked with star shaped markers. Non-EGA galaxies are marked with circle markers. EGA galaxies within Zone B are marked with magenta edges. We find that EGA galaxies typically have more concentrated SF compared to non-EGA galaxies. EGA galaxies within Zone B are typically even more concentrated than EGA galaxies found in other zones.}
    \label{vasym_c}
\end{figure}
To investigate the SF properties within EGA and non-EGA galaxies, we investigate the concentration of SF in these galaxies $C(\rm H\alpha)$ and their offset from the star-forming main sequence (SFMS). In Fig. \ref{vasym_c}, we plot $v_{\rm asym}$(Stars) against $v_{\rm asym}$ (Gas) coloured by $C(H\alpha)$. We find that the mean of $C(H\alpha)$ for EGA and non-EGA galaxies is 0.63 and 0.91, respectively, with a standard deviation of 0.20 and 0.22, respectively. While EGA galaxies appear more concentrated, there is not a statistically significant difference EGA and non-EGA galaxies. The increased concentration in EGA galaxies is consistent with gas being stripped from the outer disk from RPS, suppressing the star-formation in the outer disk. The elevated gas asymmetries is unlikely to be an artefact of the low SNR(H$\alpha$) spaxels being used to measure the gas asymmetry since galaxies with large errors in their gas asymmetry will have been excluded during our selection of EGA galaxies. For posterity, we show a plot of percentage error on the gas asymmetry against concentration in Fig. \ref{vasym_c}. All EGA galaxies have errors on the gas asymmetry that is less than 20\%, so it is unlikely to be a data quality issue causing increased concentration EGA galaxies to have elevated gas asymmetries.

\begin{figure}
    \centering
    \includegraphics[width=1\linewidth]{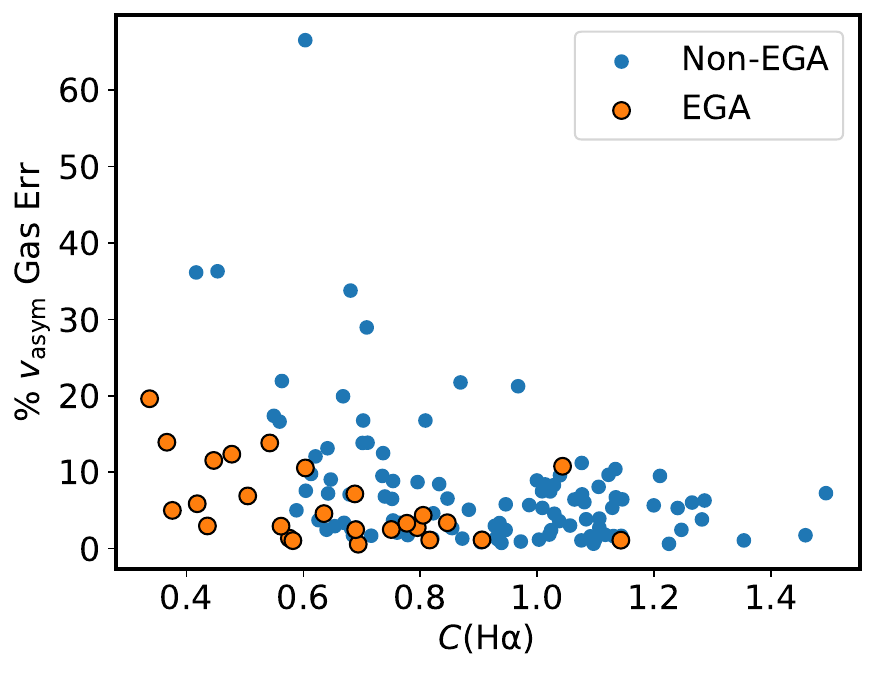}
    \caption{Gas $v_{\rm asym}$ percentage error against $C(\rm H\alpha)$. All EGA galaxies have errors on their gas asymmetry that less than 20\%, hence we can confident that it is not a data quality issue that is causing low concentration galaxies to have elevated gas asymmetries.}
    \label{vasym_c}
\end{figure}

\begin{figure}
    \centering
    \includegraphics[width=\linewidth]{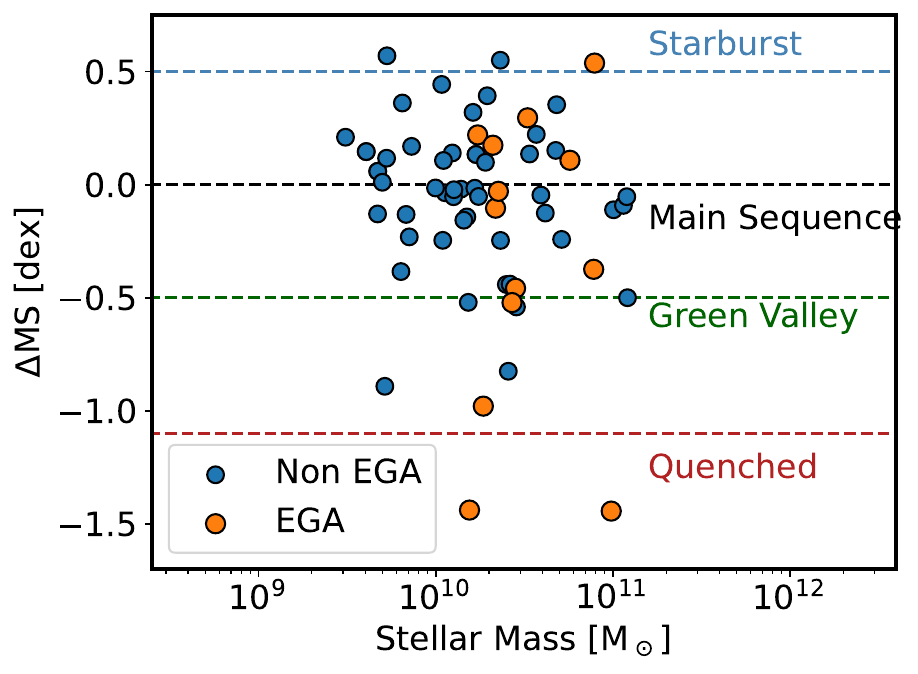}
    \caption{$\Delta$MS vs. $M_*$ for galaxies in our sample. Regions of SF activity are suggested in \citep{2020MNRAS.492...96B}. We find that most EGA galaxies are on the SFMS (8/13), with 2 galaxies in the Green Valley region, 2 galaxies in the Quenched region, and one galaxies in the Starburst region. We do not find any evidence to suggest that the \textit{global} SF in EGA galaxies is being suppressed or enhanced.}
    \label{SFMS}
\end{figure}

We also investigate the global SF properites of EGA galaxies. We quantify their offset ($\Delta$MS) from the SFMS using equation 3 from \citet{2021MNRAS.503.4992F}. To derive the SFMS, the authors used SED-derrived stellar masses and SFRs from the \textit{GALEX}-Sloan-\textit{WISE} legacy catalogue 2 \citep[GSWLC-2;][]{2016ApJS..227....2S,2018ApJ...859...11S}. To ensure a fair comparison, and avoid issues with underestimation of SFRs using H$\alpha$ fluxes in SAMI \citep[e.g.,][]{2017MNRAS.464..121S}, we also use the SED-derived SFRs and $M_*$ from the GSWLC-2, where we have matched sky coordinates between catalogues with maximum seperation of 2''. There is only coverage in GSWLC-2 for half of the clusters in our sample, with 76 galaxies in total, being in both catalogues, hence we restrict our discussion of $\Delta$MS to those 76 galaxies. Fig. \ref{SFMS} shows $\Delta$MS vs. $M_*$ for the 76 galaxies, with EGA galaxies shown in orange and non-EGA galaxies shown in blue, with classification suggested in \cite{2020MNRAS.492...96B}. We find that more than half of EGA galaxies (8/13) are on the SFMS, with 2 galaxies in the Green Valley region, 2 galaxies in the Quenched regions, and 1 galaxy (9016800318)\footnote{Galaxy 9016800318 has a bar feature that is visible in the Ha flux (See second panel of Fig. \ref{9016800318}). Non-axisymmetric features, like bars, can lead to increased kinematic asymmetries; meaning the increased $v_{\rm asym}$ (Gas) we see in this galaxy could be due to the bar, and not from the cluster environment. We investigate the individual mode sensitive to bar features (i.e., $k_3$) and find that it is the dominant mode in $v_{\rm asym}$ (Gas) (i.e., $k_3/(k_2+k_3+k_4+k_5)>0.25$). Hence, it could be that the bar is responsible for the large $v_{\rm asym}$ (Gas), and is likely funneling gas to the central regions, which is resulting in this galaxy having enhanced SF. 9016800318 is the only galaxy in our sample with a visible bar and $k_3$ as its largest mode, and excluding it from our sample does not alter our conclusions.} is in the Starburst region. This suggest that EGA galaxies are typically more concentrated, but do not tend to display \textit{globally} suppressed or enhanced SF.

It should be mentioned that the concentration values discussed here are based on light captured within the 15'' FOV of the SAMI instruments. Some EGA galaxies may have both $r-$band and H$\alpha$ emission that extends beyond the FOV of SAMI instruments, meaning the $C(\mathrm{H}\alpha)$ used in this work may not be a true indication of how concentrated the star-formation is. We check if some EGA galaxies possibly have underestimated concentration by comparing the SDSS $r-$band $R_\mathrm{e}$ with the SAMI FOV and we find that all EGA galaxies barring three (9011900599, 9388001017 and 9239900205) have $r-$band $R_\mathrm{e}$ less than 7.5'' (i.e., SAMI FOV/2). When visually inspecting these three galaxies, we see that H$\alpha$ emission appears to extend beyond the SAMI FOV. This means that the concentration measured in SAMI is likely not accurate and we cannot confirm that it does indeed have increased concentration. For galaxies with an SDSS $r-$band $R_\mathrm{e}$ comfortably within the SAMI FOV (23/26 of EGA galaxies), we can be confident the concentration is accurate and that these 22 galaxies are, on average, more concentrated than non-EGA galaxies; however, we leave this discussion behind the concentration of star-formation as a caveat of our analysis. Continuum and H$\alpha$ flux, maps as well as stellar and gas velocity maps for the EGA galaxies are shown in App. \ref{flux_velo_images} with 9011900599, 9403800169 and 9239900205 shown with red axis edges.

\begin{figure*}
    \centering
    \includegraphics[width=0.90\linewidth]{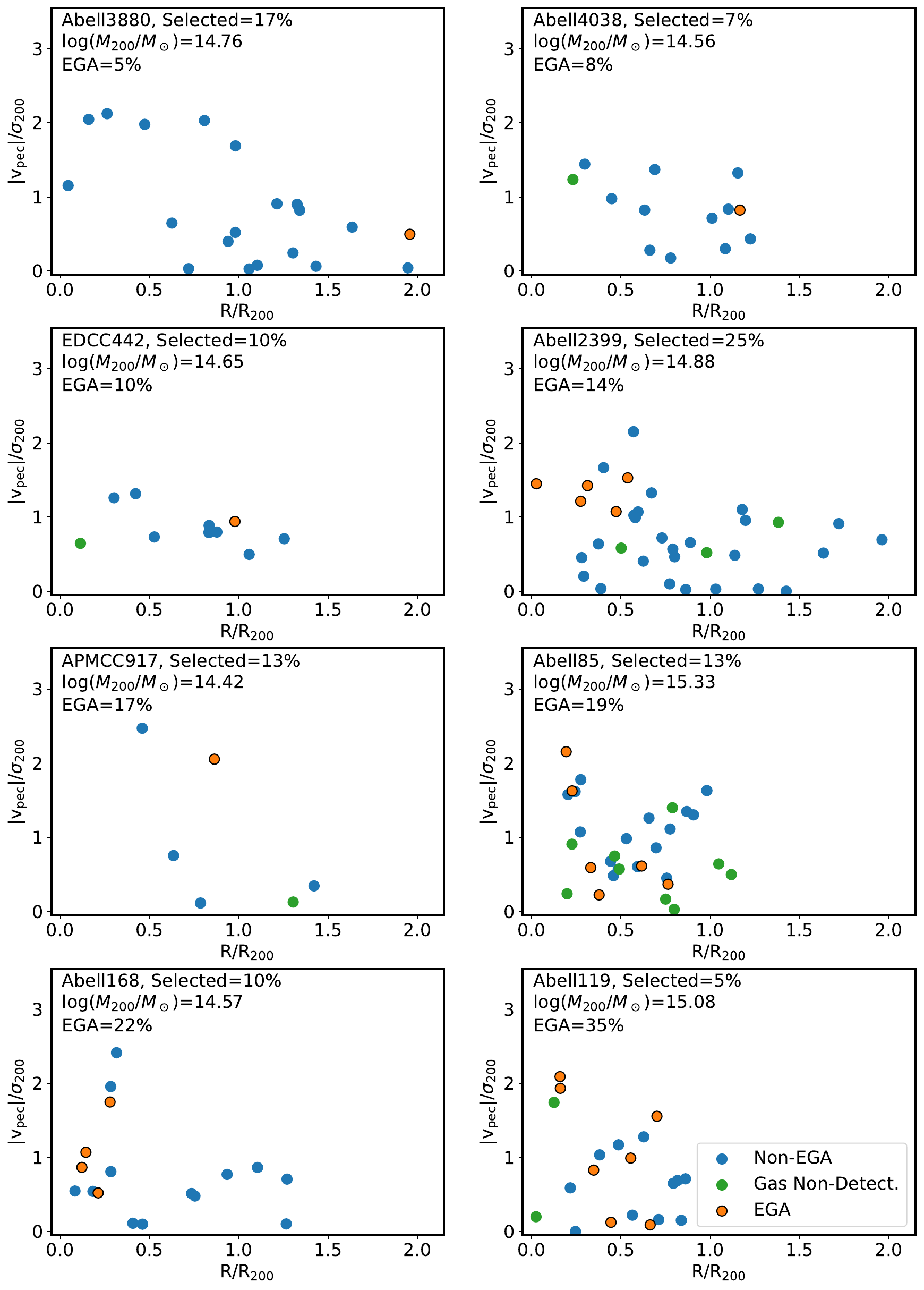}
    \caption{PPS for each cluster in our sample with increasing fraction of EGA galaxies within the cluster. EGA galaxies are shown in orange, non-EGA galaxies are shown in blue and gas non-detections are shown in green. The fraction of galaxies selected from each cluster is shown next to the name of the cluster. The fraction of EGA galaxies tends to increase with $M_{200}$ and with the prevalence of substructure within the cluster.}
    \label{PPS_sersic_cluster}
\end{figure*}

\subsection{Dynamical Evolution due to Cluster Environment}
It is interesting that EGA galaxies are preferentially found in PPS where RPS is predicted to be most significant (i.e., Zone B). As argued above, it is likely that RPS removes the gas in the outskirts of these galaxies, kinematically disturbing the gas (only) in the process and thus causing elevated gas asymmetries. The removal of cold gas consequently suppresses SF in the outer disk and it begins to fade while the existing stellar disk or bulge survive, and naturally explains why EGA galaxies typically are more concentrated than non-EGA galaxies. Disc-fading has been associated with the dynamical transformation of galaxies from fast to slow rotating galaxies, where the more dispersion supported bulge begins to dominate the mass within the galaxy \citep[e.g.,][]{2021MNRAS.505..991C}, and is typically invoked to explain the increased number of lenticular, and decreased number of spiral galaxies with increasing environmental density.

It should be noted that EGA galaxies do not typically display either globally enhanced or suppressed SF (See Sect. \ref{SFA}), despite these galaxies being typically more concentrated. This suggests that some EGA galaxies have yet to, or are beginning to, undergo quenching globally while they experience the peak of RPS in clusters, but are still undergoing `outside-in' quenching from the stripping of cold gas in their outskirts. This is consistent with the scenario of gas stripping happening quickly, particularly at the peak of RPS, but quenching acting over longer timescales \citep[i.e., delayed-then-rapid quenching;][]{2016MNRAS.463.3083O,2020ApJS..247...45R}. Also, all EGA galaxies in our sample are all massive ($M_*>10^{10}$ $M_\odot$) galaxies, which can still retain significant reservoirs of cold gas in their central regions, even after first pericentre passage shown by their gas depletion times ($M_{\rm gas}$/SFR) exceeding billions of years \citep[e.g.,][]{2009MNRAS.400.1225C,2014A&A...570A..69B,2021PASA...38...35C}.

Finding residual SF in the inner regions of galaxies is common for galaxies undergoing environmental quenching since it is unclear whether environmental effects are strong enough to impact the inner regions. \cite{2023PASA...40...17W} combined spatially resolved atomic Hydrogen (\HI) and CO observations to investigate the impact of environment on molecular Hydrogen ($H_2$) compared to \HI. Quantifying the effect of environment by how \HI\ deficient galaxies are, the authors found that \HI\ within the central regions is as effectively stripped as the outer regions, removing the densest regions of \HI. Conversely, for $H_2$, when the environmental impact is most severe, the densest molecular gas regions can survive, but the molecular gas surface density of the whole disk decreases. This reduction in average molecular gas surface density likely results from the molecular gas being redistributed from the central regions lowering the average surface density. Since molecular gas surface density is tightly coupled to SFR surface density \citep[e.g.,][]{1959ApJ...129..243S,1989ApJ...344..685K,2008AJ....136.2782L,2022ARA&A..60..319S} the spatial redistribution through environmental stripping could be responsible for the elevated gas asymmetries in the ionised gas. 

Similarly, \cite{2021MNRAS.500.1285B} investigated the effects of RPS on `unwinding' spiral arms for spiral galaxies within a cluster. They find that the pitch angle (i.e., the angle between spiral arms) increases radially in all spiral galaxies in their sample, and this primarily affects the younger stellar populations, leaving the older stellar populations in place. Using idealized hydrodynamical simulations, the authors find this `unwinding', when the galaxy is face-on with respect to the direction of infall, arises from gas within the spiral arms moving to orbits at larger radii. This is consistent with our results where the younger stellar populations, traced by the ionised gas, becoming decoupled from the existing stellar disk through RPS, which would lead to larger gas asymmetries. 

Fig. \ref{PPS_sersic_cluster} shows the PPS for each cluster in our sample with increasing fraction of EGA galaxies within the cluster. We find that the fraction of EGA galaxies generally increases with $M_{200}$, however it is not statistically significant correlation ($\rho_{\rm spearman}$=0.46,p=0.25). The fraction of EGA galaxies also increases with the presence of substructure within the cluster. We do not find a trend between fraction of EGA galaxies in the clusters and the fraction of galaxies sampled from individual clusters, suggesting the number of EGAs is not related to how many galaxies are within a cluster. The substructure within Abell 85, Abell 168 and Abell 2399 is well documented from X-ray and optical observations \citep{2002ApJ...579..236K,2007A&A...469..363B,2014MNRAS.443..485F,2016ApJ...832...69O,2023ApJ...955..103W}, however, \cite{2017MNRAS.468.1824O} found no strong evidence of substructure within APMCC917, Abell 3880, Abell 4038 or EDCC442 and marginal evidence of substructure in Abell119, but it is suggested to still be a dynamically complex cluster \citep[e.g.,][]{2016ApJ...822...92L}. The excess of EGA galaxies could be explained by the larger velocities galaxies will have in merging clusters compared to more dynamically relaxed clusters, and the larger relative velocities will lead to enhanced RPS. The prevalence of EGA galaxies in clusters with significant substructure could also suggest that EGA galaxies belong to galaxy groups that have fallen into, but not yet fully diffused into the cluster. Finally, the increasing fraction of EGA galaxies in higher $M_{200}$ clusters could also be related to mass of the cluster itself, rather than substructure or previous group environments; where higher mass cluster will have a denser ICM, resulting in stronger RPS during pericentre passage. Disentangling the individual contribution of halo mass and substructure will require a selecting EGA galaxies from clusters with a narrow range of halo masses, which is beyond the scope of this paper.

So far, we have been assuming that EGA galaxies have not undergone significant evolution within the group environment before entering the clusters \citep[i.e., pre-processing,][]{2004PASJ...56...29F,2012MNRAS.423.1277D,2017ApJ...843..128R,2022MNRAS.511.3210P}. Galaxies being pre-processed could explain the number of gas non-detections we find in our sample, particularly in the clusters shown to have significant substructure (Abell2399, Abell85 and Abell119). Quenching for massive galaxies being pre-processed happens on short timescales \cite[e.g.,][]{2019MNRAS.486..868K,2021PASA...38...35C,2024MNRAS.529.3651O}, meaning if the EGA galaxies had their star-formation pre-processed before entering the cluster, they would have been quenched long before they entered the cluster. Similar to star-formation pre-processing, \citep{2025arXiv250108461F} explored how environment effects galaxy dynamics by searching for the kinematic morphology-density relation in the Middle Ages Galaxy Properties in IFS Survey \citep[MAGPI;][]{2021PASA...38...31F}, and found a strong tension between MAGPI and SAMI galaxies, mainly that slow or non-obvious rotating MAGPI galaxies were not preferentially found in higher $M_{200}$ halos. The authors suggested that MAGPI galaxies were being `dynamically pre-processed' in smaller halos before eventually migrating to higher-density environments. EGA galaxies being dynamically pre-processed in group environments first, and may also explain the high fraction of EGA galaxies in clusters with significant substructures.

\section{Conclusion}
\label{conc}
We have conducted a study of the dynamical evolution of galaxies as they fall into cluster environments using a sample of galaxies from the SAMI Galaxy Cluster Survey. Using \textsc{kinemetry} to model the kinematics of stars and gas, we measure the kinematic asymmetry present in the stellar and ionised gas velocity fields to gauge how the cluster environment disturbs the stellar and ionised gas kinematics. By comparing the stellar and gas kinematic asymmetries, we seek to separate the hydrodynamical (e.g., ram pressure stripping, viscous stripping, evaporation etc.) and gravitational (e.g., interactions) processes that occur in clusters to constrain which are driving their dynamical evolution. To do this, we select galaxies where their gas asymmetries are significantly elevated above their stellar asymmetries (e.g., EGA galaxies).

Our findings are as follows:
\begin{itemize}
    \item We find that the fraction of EGA galaxies is $17^{^{+2}}_{-3}\%$ (significantly above the fraction of EGAs found in non-cluster environment). Examining the Projected-Phase Space (PPS) for galaxies within our sample, we find the highest fractions of EGA galaxies  within the `recent infaller' zone (38\%) and `ancient infaller' zone (30\%, Fig. \ref{PPS_infall}). After correcting the fractions of EGA galaxies in the `ancient infaller' zone for potential contamination of recent infallers, we find that it is consistent with the fraction of EGA galaxies found in non-cluster environments. Galaxies within the `recent infaller' zone are expected to experience the peak of ram pressure stripping (RPS), and the significantly higher fraction of galaxies in this zone points to RPS being exclusively responsible.
    \item To isolate the effect of the cluster environment, we plot $v_{\rm asym}$(Gas)/$v_{\rm asym}$(Stars) against $5^{\rm th}$-nearest-neighbour density ($\Sigma_5$) for cluster and non-cluster galaxies. At the same $\Sigma_5$, cluster galaxies have a larger $v_{\rm asym}$(Gas)/$v_{\rm asym}$(Stars) however it is not a significant difference. After matching stellar mass and redshift distributions, the difference becomes even less significant.
    \item We find that EGA galaxies typically have more concentrated SF than non-EGA galaxies, with an average $C(\rm H\alpha)$ of 0.63$\pm0.20$ and 0.91$\pm 0.22$, respectively. This could possibly indicate outside-in quenching and disk-fading occurring in EGA galaxies, most likely due to RPS.
    \item Finally, we find that clusters with the largest fraction of EGA galaxies are those with significant substructure (Fig. \ref{PPS_sersic_cluster}). We suggest that EGA galaxies belong to groups that have recently fallen into the cluster, and are primarily undergoing disc-fading due to gas stripping from hydrodynamical processes exclusive to clusters (i.e., RPS), based on their position in PPS and concentration.
\end{itemize}

In this work, we have demonstrated that stellar and ionised gas kinematic asymmetries in galaxy clusters can be used to disentangle various cluster-specific physical processes. While this work considered only the stellar and ionised gas kinematics, the analysis may be extended to other gas phases with 2D kinematic maps that can be modeled in a similar fashion (e.g., \HI\ and $H_2$). Comparing different baryon phases would allow a further decomposition of hydrodynamical effects in cluster environments, since \HI\ is more affected by environment than $H_2$ \citep[e.g.,][]{2014A&A...564A..66B,2021MNRAS.502.3158S} and acts on different spatial scales with environmental impact still measurable on high molecular surface densities \citep[e.g.,][]{2023PASA...40...17W}. Multiwavelength surveys of cluster environments, like the Virgo Environment Traced in CO Survey \citep[VERTICO;][]{2021ApJS..257...21B} and MUSE and ALMA Unveiling the Virgo Environment \citep[MAUVE;][]{2024MNRAS.530.1968W}, are expected to further our understanding of how the cluster environment affects individual gas phases.

\section*{Acknowledgements}
RSB would like to thank Aman Khalid for his help plotting the discrete regions in PPS. We also thank the referee for a constructive referee report.

The SAMI Galaxy Survey is based on observations made at the Anglo-Australian Telescope. The Sydney-AAO Multi-object Integral field spectrograph (SAMI) was developed jointly by the University of Sydney and the Australian Astronomical Observatory. The SAMI input catalogue is based on data taken from the Sloan Digital Sky Survey, the GAMA Survey and the VST ATLAS Survey. The SAMI Galaxy Survey was supported by the Australian Research Council Centre of Excellence for All Sky Astrophysics in 3 Dimensions (ASTRO 3D), through project number CE170100013, the Australian Research Council Centre of Excellence for All-sky Astrophysics (CAASTRO), through project number CE110001020, and other participating institutions. The SAMI Galaxy Survey website is http://sami-survey.org/

CF is the recipient of an Australian Research Council Future Fellowship (project number FT210100168) and Discovery Project (project number DP210101945) funded by the Australian Government. OÇ acknowledges the financial support from the Australian Government Research Training Program Scholarship (RTP).

\section*{Data Availability}
The SAMI Galaxy Survey website is \url{http://sami-survey.org/}, and all data used in this work is publicly available through Data Central \url{https://datacentral.org.au}.



\bibliographystyle{mnras}
\bibliography{reference} 



\appendix
\section{Flux and velocity maps for Elevated Gas Asymmetry Galaxies}
\label{flux_velo_images}
\begin{figure*}
    \centering
    \includegraphics[width=1\linewidth]{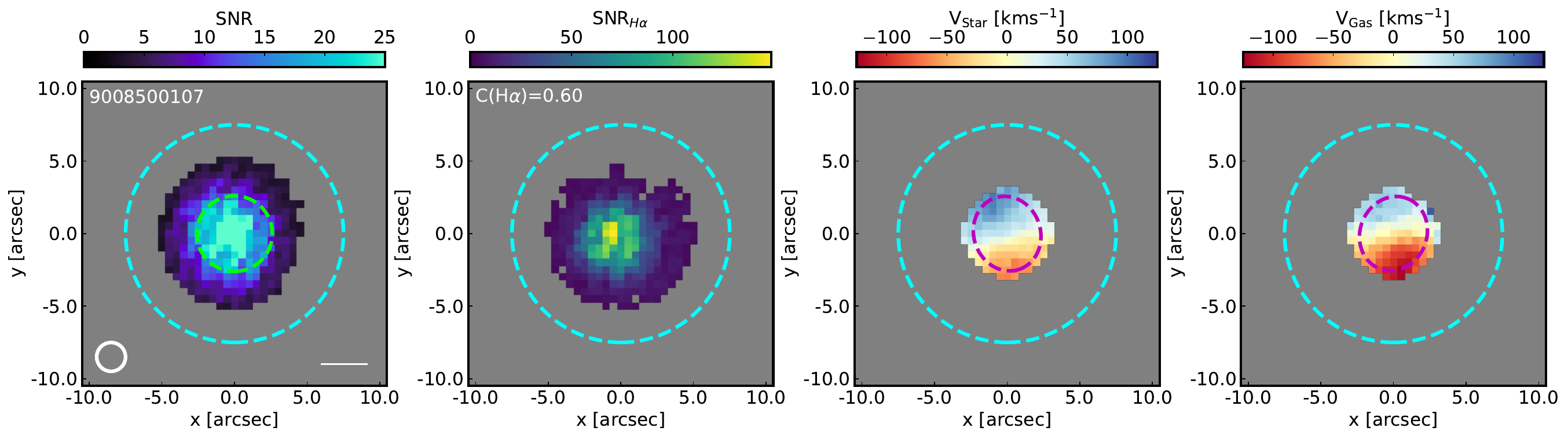}
    \caption{Continuum flux map (\textit{left}), H$\alpha$ flux map (\textit{middle left}), stellar velocity (\textit{middle right}) and gas velocity maps (\textit{right} for galaxy 9008500107. Spaxels with SNR<3 within a 2$R_e$ circular aperture are masked for the flux maps, and spaxels within 1$R_e$ are masked for the velocity maps. A larger masking aperture is used for the flux maps to show extended emission outside 1$R_e$. Magenta ellipse represent the $R_e$ for each galaxy where the kinematic asymmetry is measured. The cyan circle represents the SAMI FOV. The PSF for the observation is shown in the lower left and the horizontal line represents a 10 kpc scale.}
    \label{9008500107}
\end{figure*}

\begin{figure*}
    \centering
    \includegraphics[width=1\linewidth]{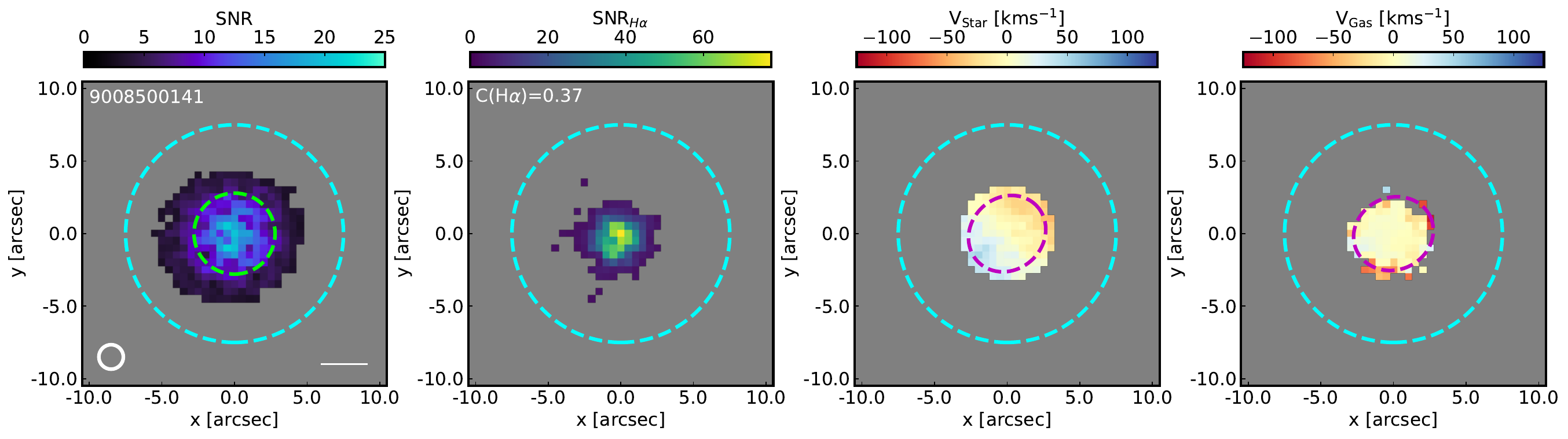}
    \caption{The same as Fig. \ref{9008500107} but for galaxy 9008500141}
    \label{9008500141}
\end{figure*}

\begin{figure*}
    \centering
    \includegraphics[width=1\linewidth]{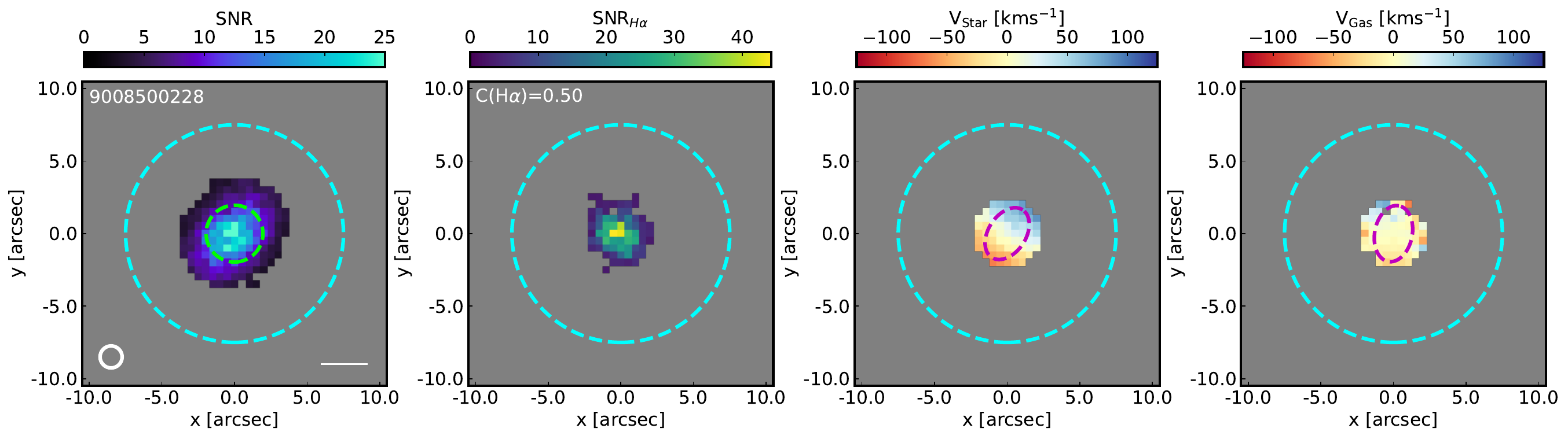}
    \caption{The same as Fig. \ref{9008500107} but for galaxy 9008500228}
    \label{9008500228}
\end{figure*}

\begin{figure*}
    \centering
    \includegraphics[width=1\linewidth]{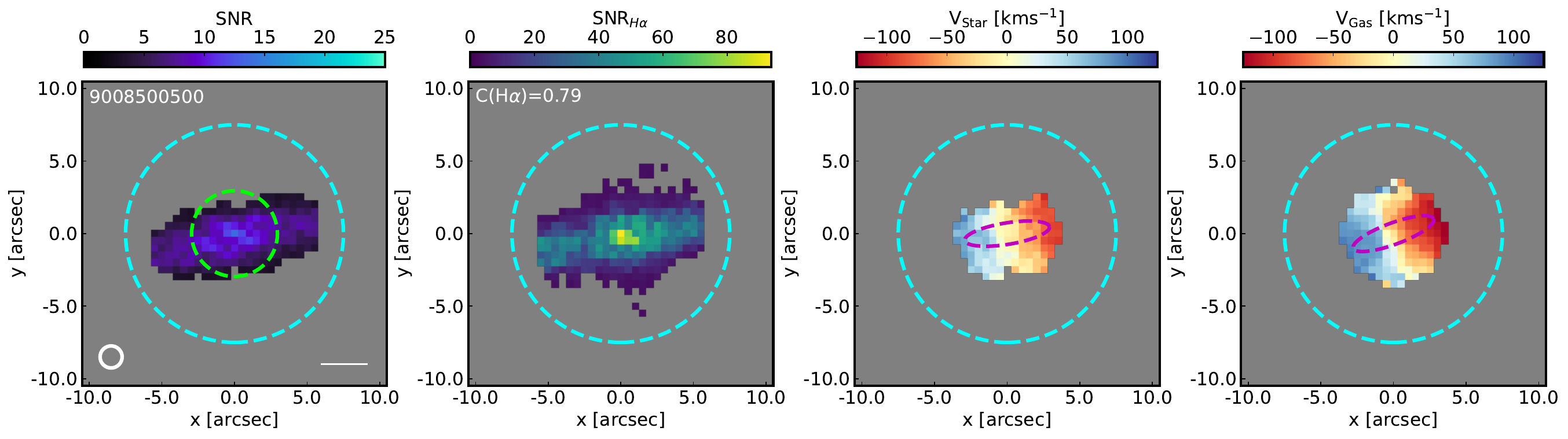}
    \caption{The same as Fig. \ref{9008500107} but for galaxy 9008500500}
    \label{9008500500}
\end{figure*}

\begin{figure*}
    \centering
    \includegraphics[width=1\linewidth]{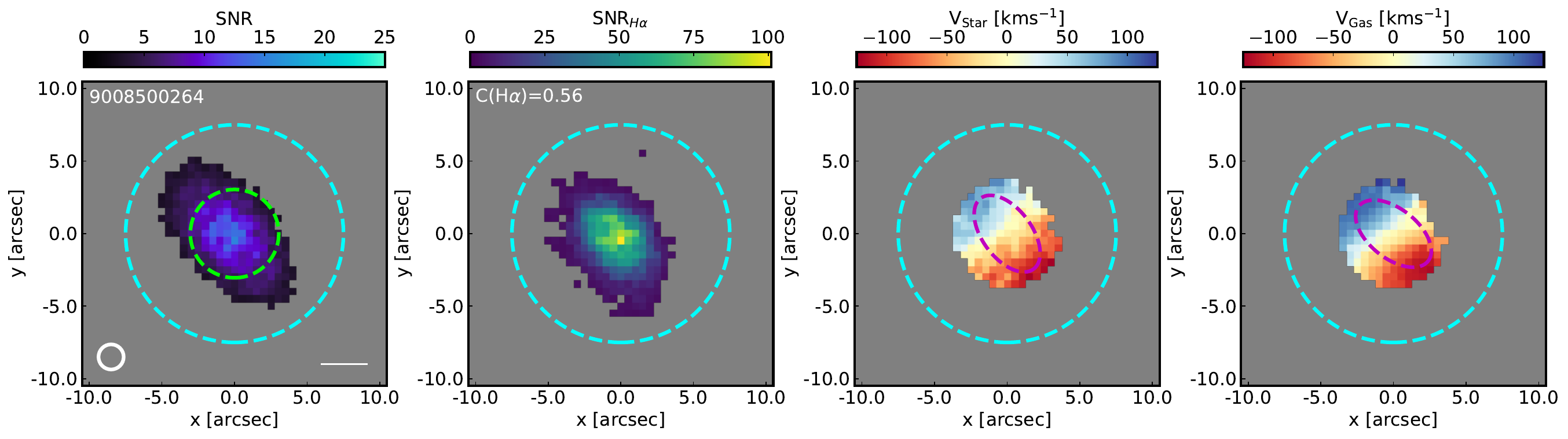}
    \caption{The same as Fig. \ref{9008500107} but for galaxy 9008500264}
    \label{9008500690}
\end{figure*}

\begin{figure*}
    \centering
    \includegraphics[width=1\linewidth]{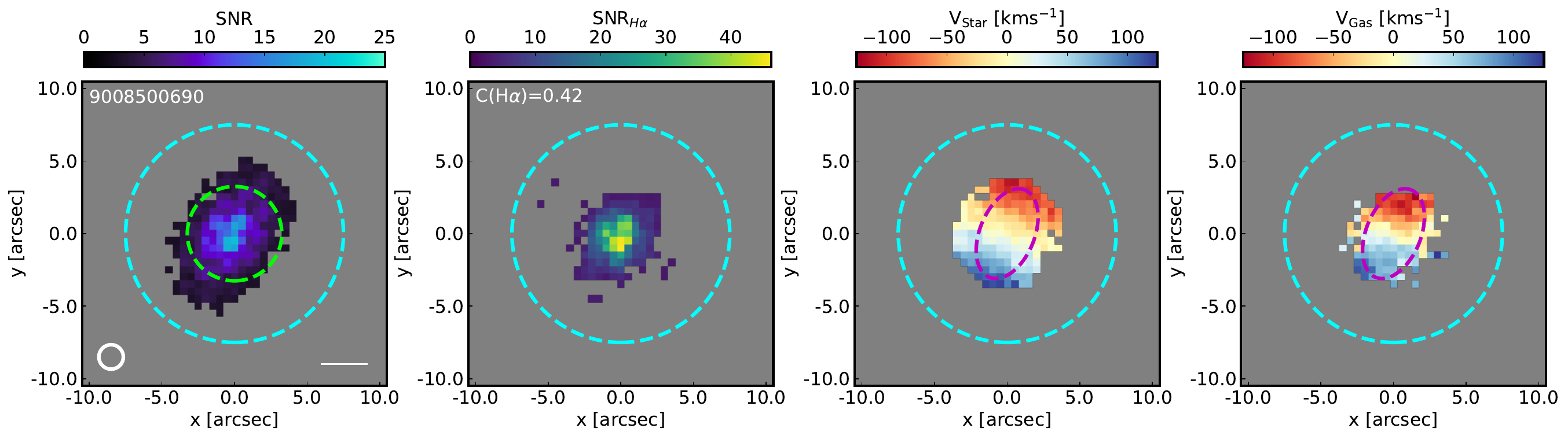}
    \caption{The same as Fig. \ref{9008500107} but for galaxy 9008500690}
    \label{9008500690}
\end{figure*}

\begin{figure*}
    \centering
    \includegraphics[width=1\linewidth]{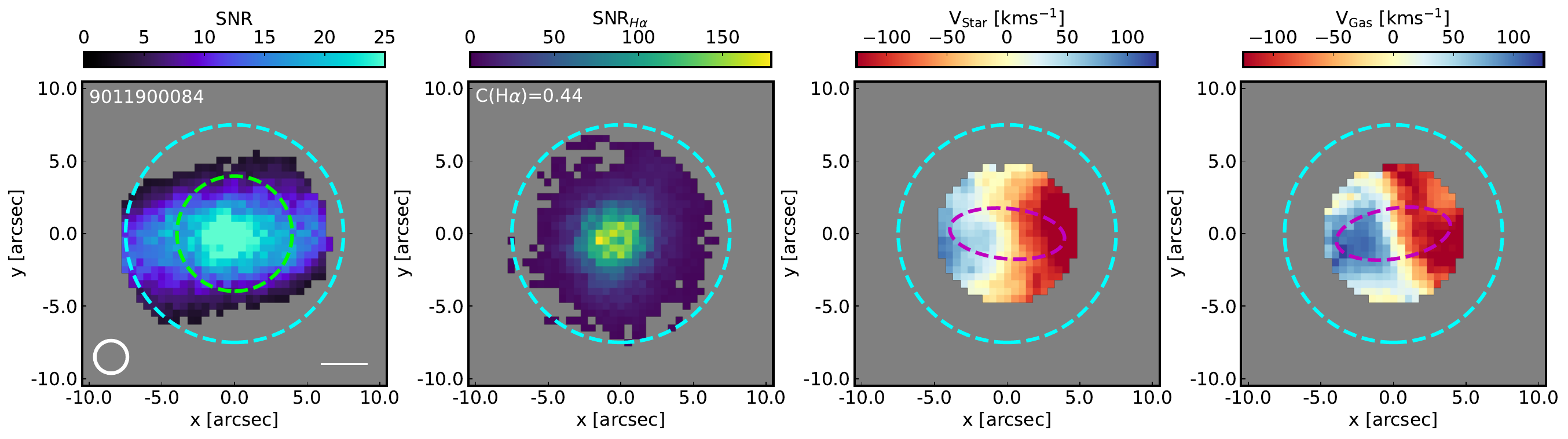}
    \caption{The same as Fig. \ref{9008500107} but for galaxy 9011900084}
    \label{9011900084}
\end{figure*}

\begin{figure*}
    \centering
    \includegraphics[width=1\linewidth]{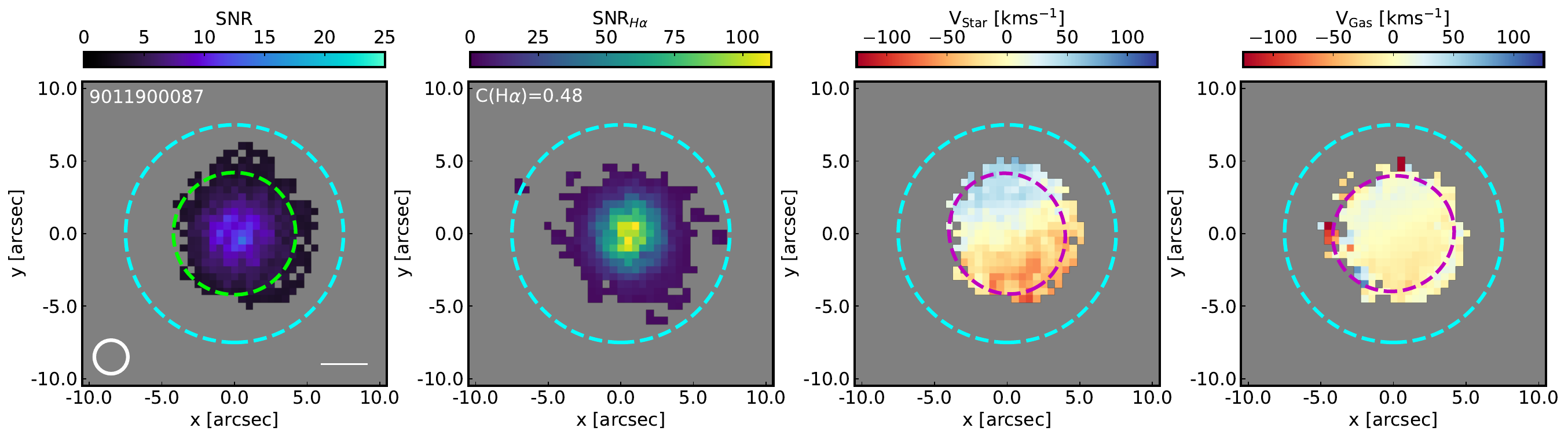}
    \caption{The same as Fig. \ref{9008500107} but for galaxy 9011900087}
    \label{9011900087}
\end{figure*}

\begin{figure*}
    \centering
    \includegraphics[width=1\linewidth]{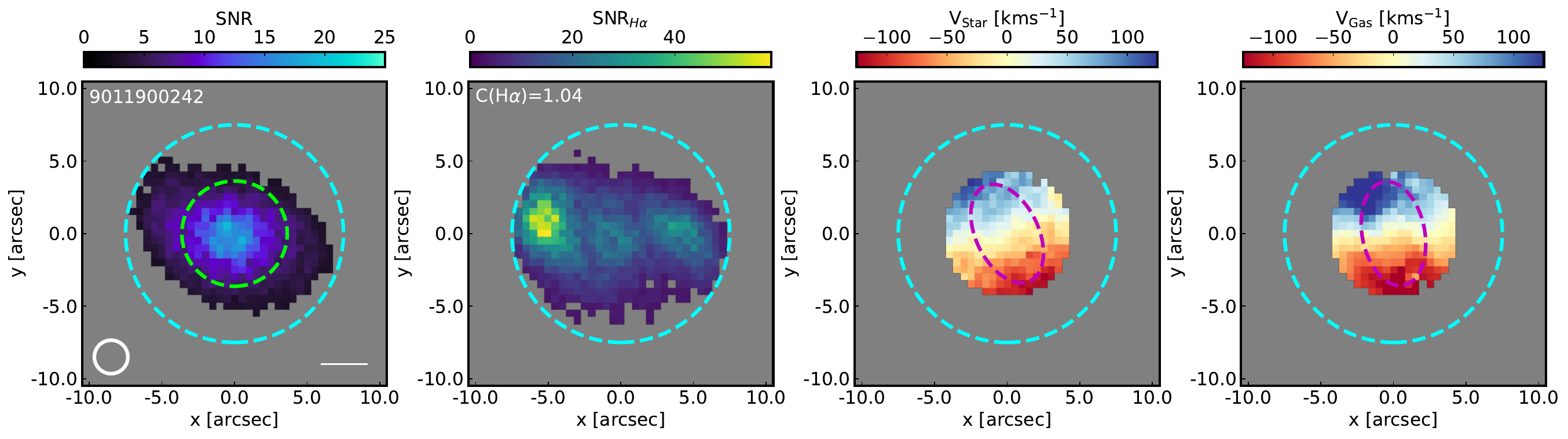}
    \caption{The same as Fig. \ref{9008500107} but for galaxy 9011900242}
    \label{9011900242}
\end{figure*}

\begin{figure*}
    \centering
    \includegraphics[width=1\linewidth]{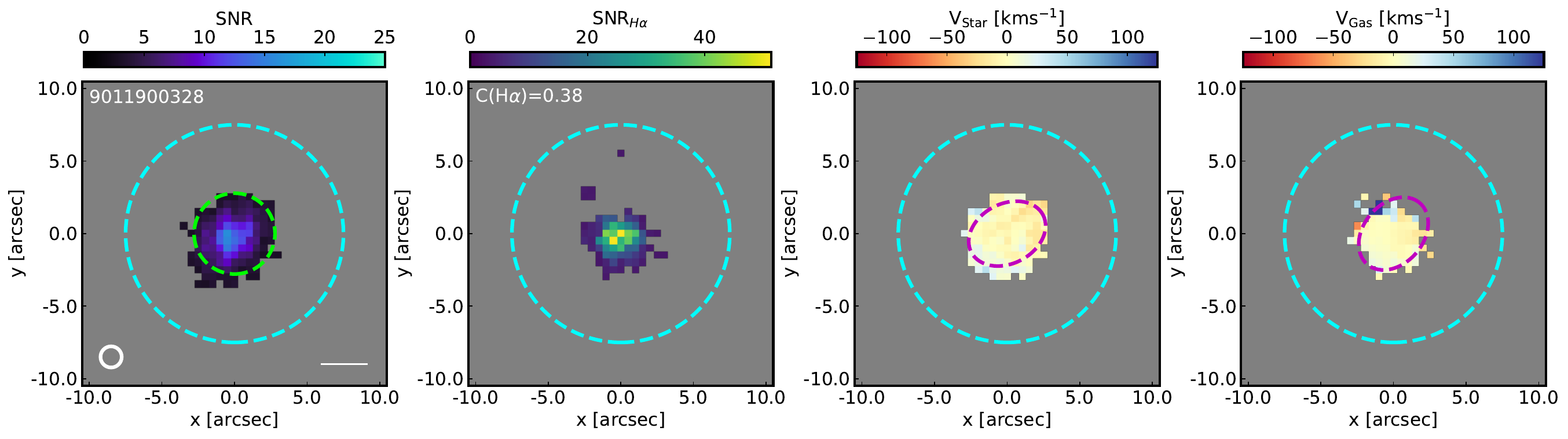}
    \caption{The same as Fig. \ref{9008500107} but for galaxy 9011900328}
    \label{9011900328}
\end{figure*}

\begin{figure*}
    \centering
    \includegraphics[width=1\linewidth]{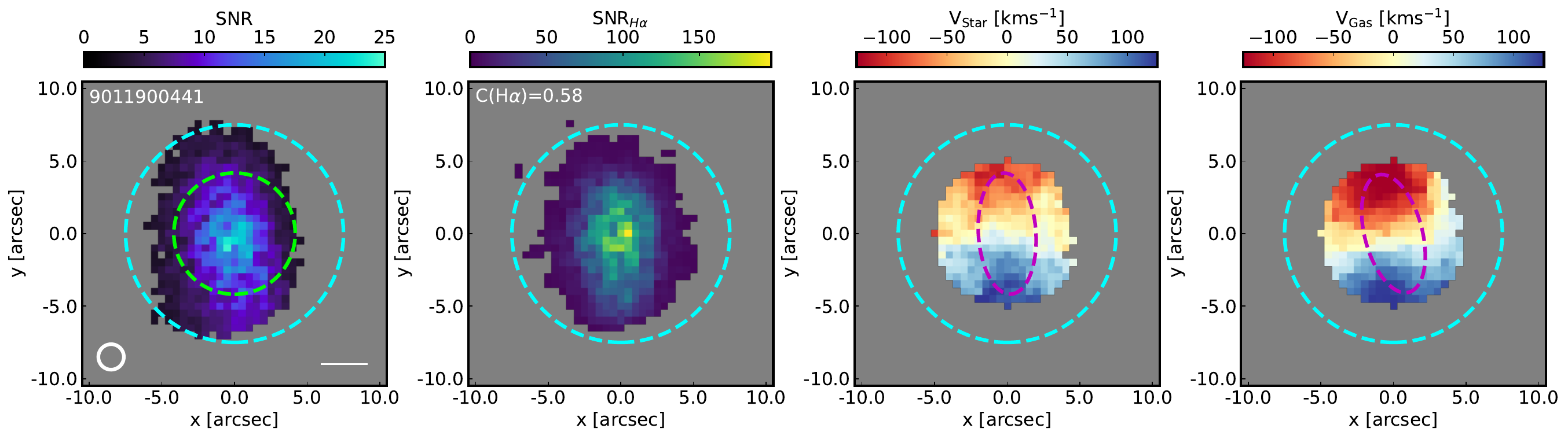}
    \caption{The same as Fig. \ref{9008500107} but for galaxy 9011900441}
    \label{9011900441}
\end{figure*}

\begin{figure*}
    \centering
    \includegraphics[width=1\linewidth]{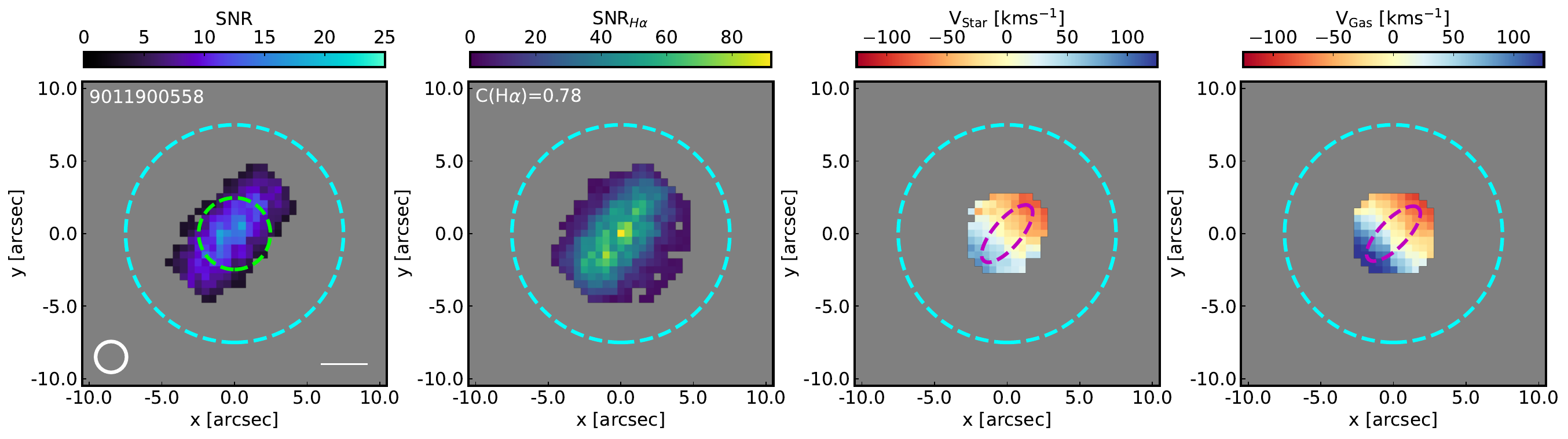}
    \caption{The same as Fig. \ref{9008500107} but for galaxy 9011900588}
    \label{9011900588}
\end{figure*}

\begin{figure*}
    \centering
    \includegraphics[width=1\linewidth]{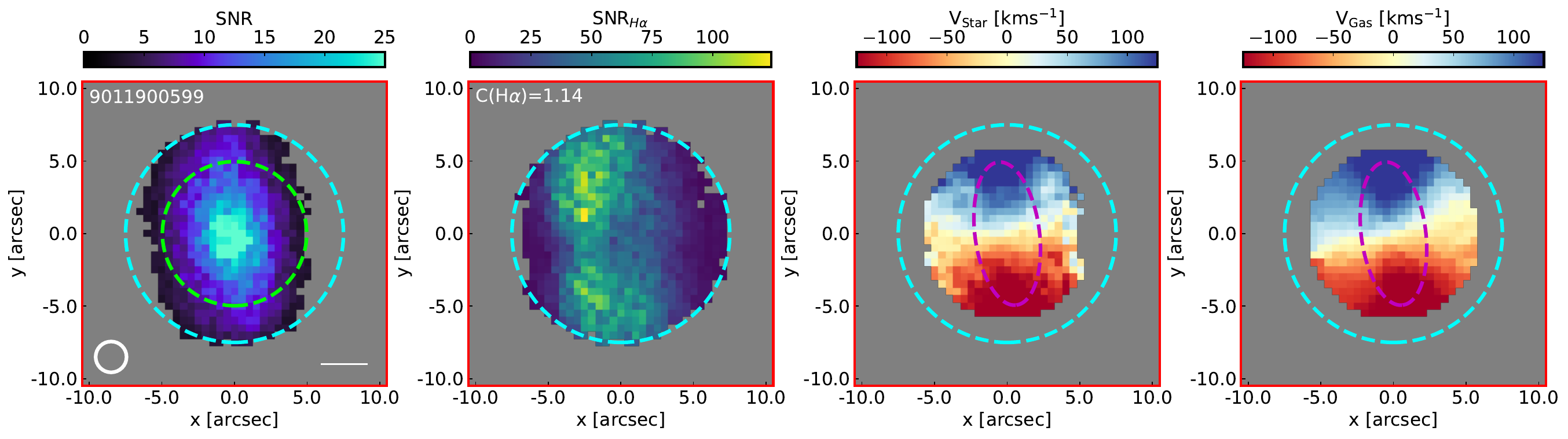}
    \caption{The same as Fig. \ref{9008500107} but for galaxy 9011900599}
    \label{9011900599}
\end{figure*}

\begin{figure*}
    \centering
    \includegraphics[width=1\linewidth]{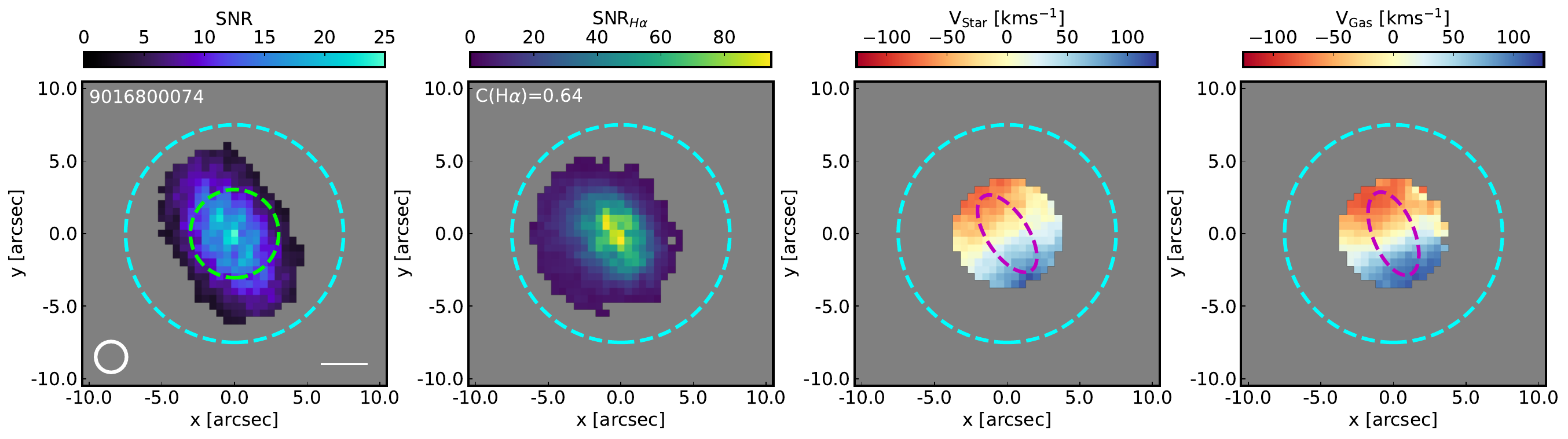}
    \caption{The same as Fig. \ref{9008500107} but for galaxy 9016800074}
    \label{9016800110}
\end{figure*}

\begin{figure*}
    \centering
    \includegraphics[width=1\linewidth]{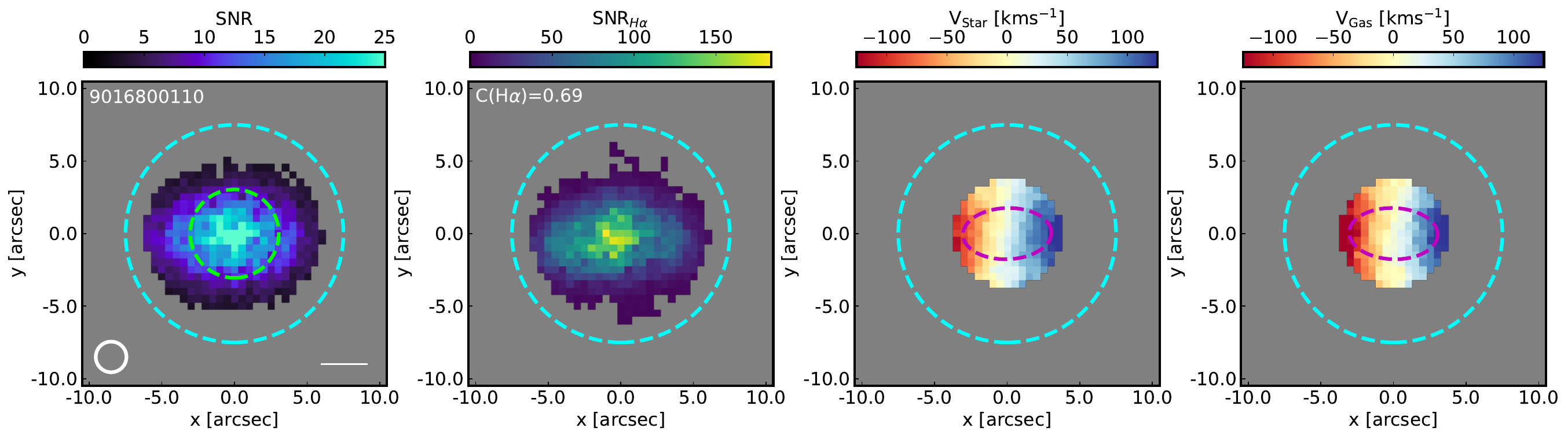}
    \caption{The same as Fig. \ref{9008500107} but for galaxy 9016800110}
    \label{9016800110}
\end{figure*}

\begin{figure*}
    \centering
    \includegraphics[width=1\linewidth]{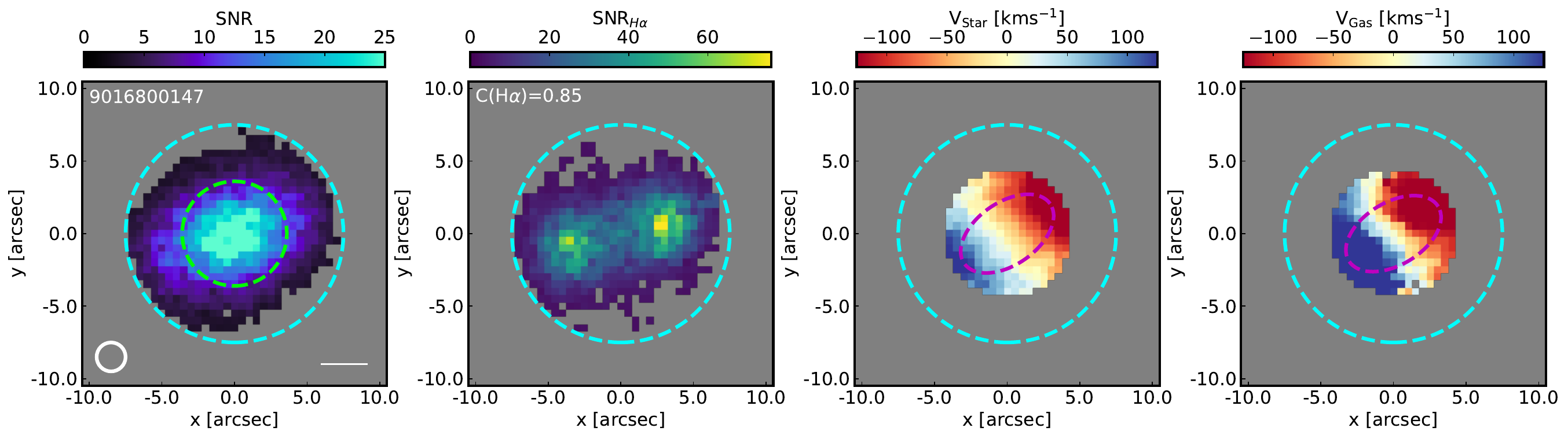}
    \caption{The same as Fig. \ref{9008500107} but for galaxy 9016800147}
    \label{9016800110}
\end{figure*}

\begin{figure*}
    \centering
    \includegraphics[width=1\linewidth]{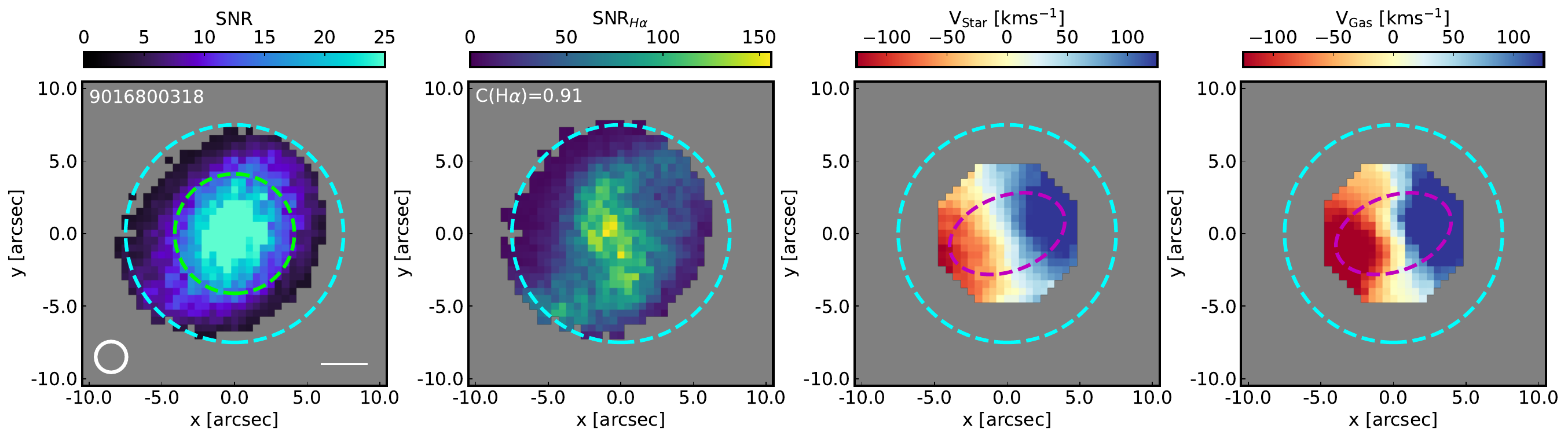}
    \caption{The same as Fig. \ref{9008500107} but for galaxy 9016800318}
    \label{9016800318}
\end{figure*}

\begin{figure*}
    \centering
    \includegraphics[width=1\linewidth]{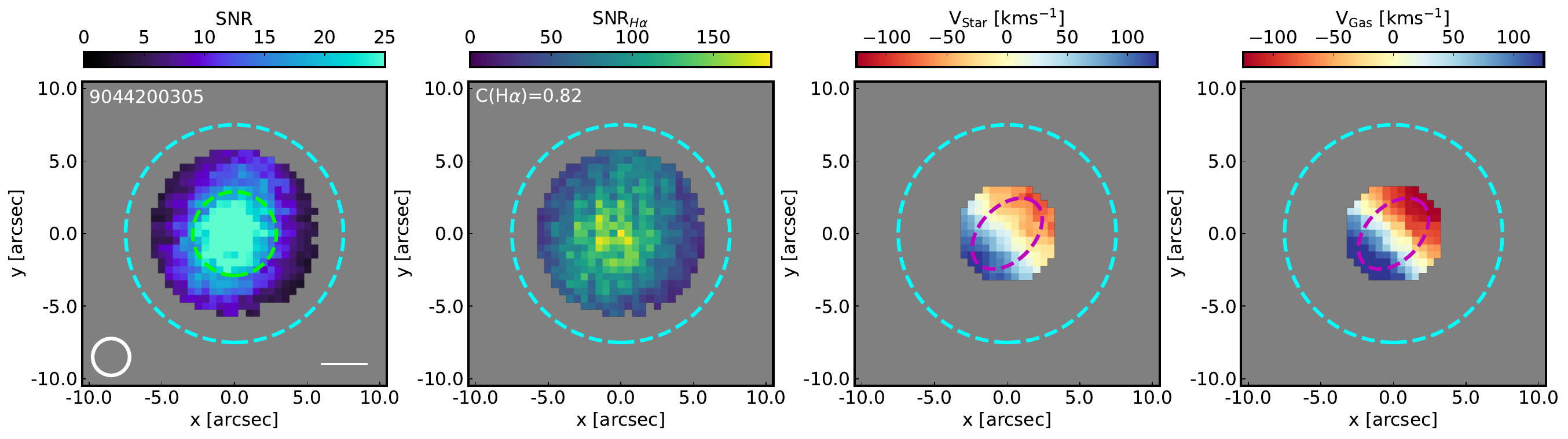}
    \caption{The same as Fig. \ref{9008500107} but for galaxy 9044200305}
    \label{9044200305}
\end{figure*}

\begin{figure*}
    \centering
    \includegraphics[width=1\linewidth]{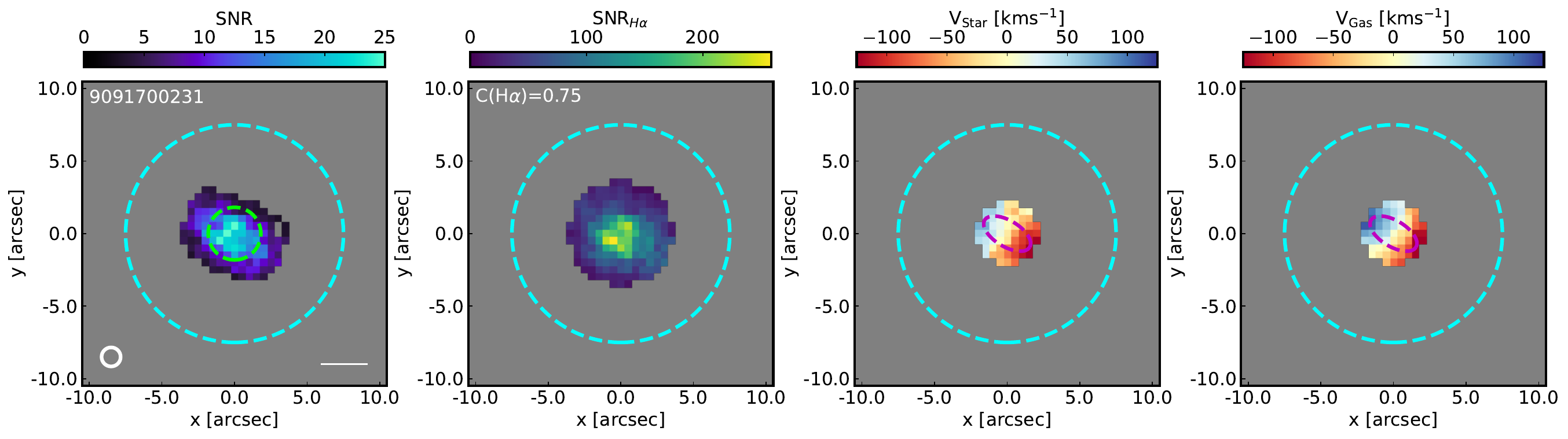}
    \caption{The same as Fig. \ref{9008500107} but for galaxy 9091700231}
    \label{9091700231}
\end{figure*}

\begin{figure*}
    \centering
    \includegraphics[width=1\linewidth]{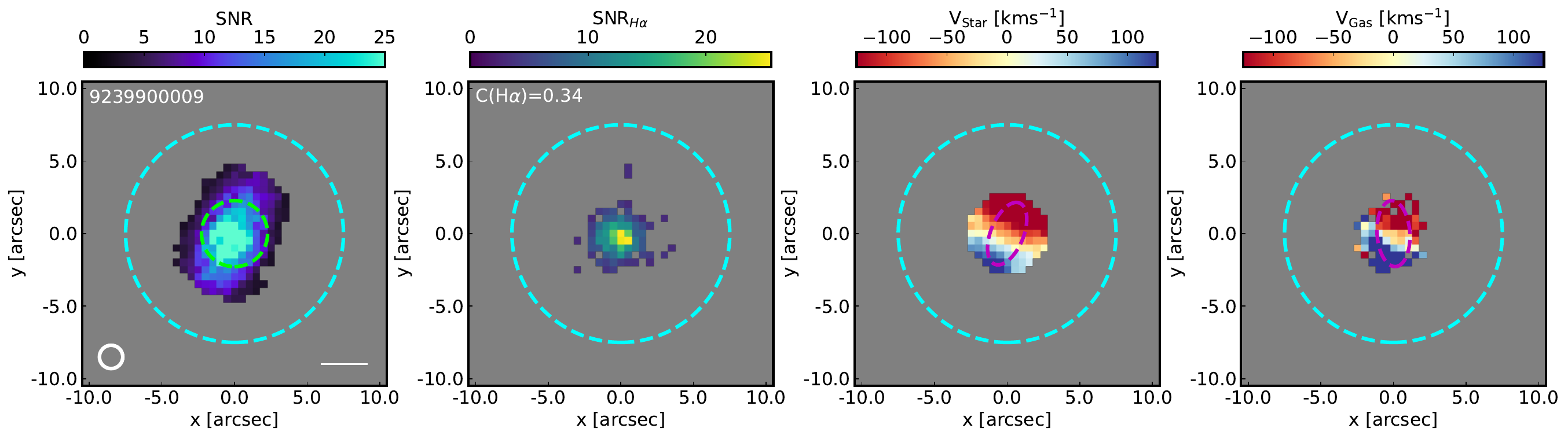}
    \caption{The same as Fig. \ref{9008500107} but for galaxy 9239900009}
    \label{9239900009}
\end{figure*}

\begin{figure*}
    \centering
    \includegraphics[width=1\linewidth]{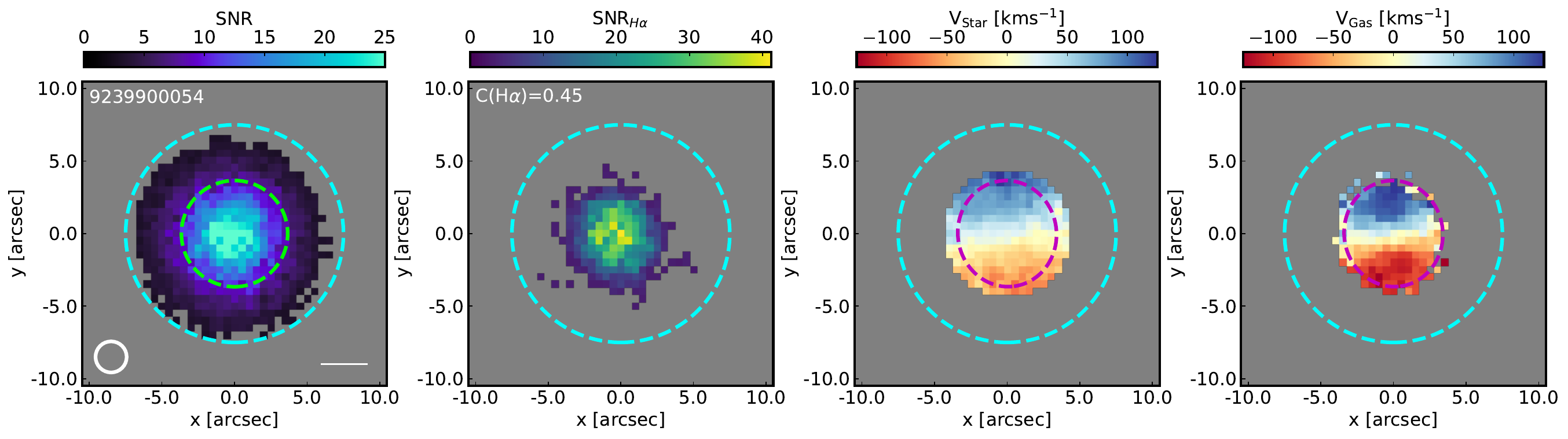}
    \caption{The same as Fig. \ref{9008500107} but for galaxy 9239900054}
    \label{9239900054}
\end{figure*}

\begin{figure*}
    \centering
    \includegraphics[width=1\linewidth]{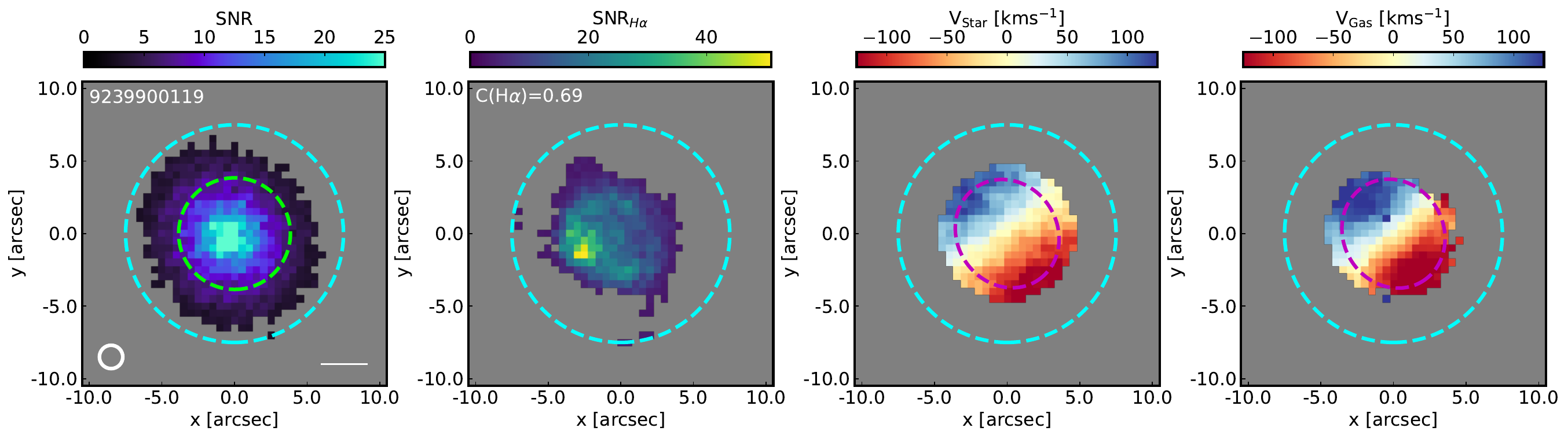}
    \caption{The same as Fig. \ref{9008500107} but for galaxy 9239900119}
    \label{9239900182}
\end{figure*}

\begin{figure*}
    \centering
    \includegraphics[width=1\linewidth]{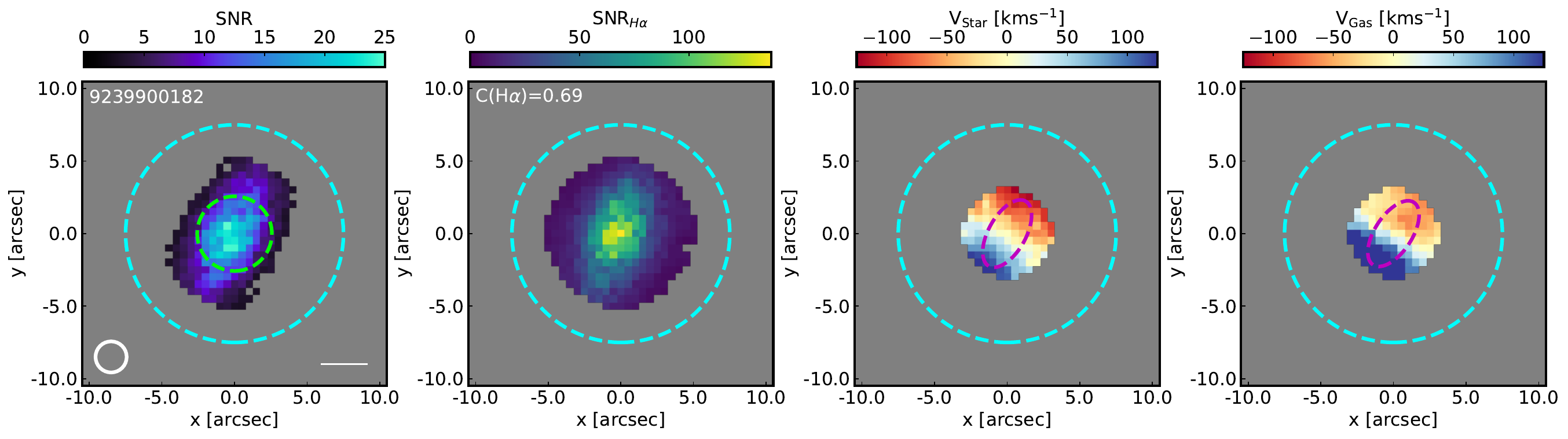}
    \caption{The same as Fig. \ref{9008500107} but for galaxy 9239900182}
    \label{9239900182}
\end{figure*}

\begin{figure*}
    \centering
    \includegraphics[width=1\linewidth]{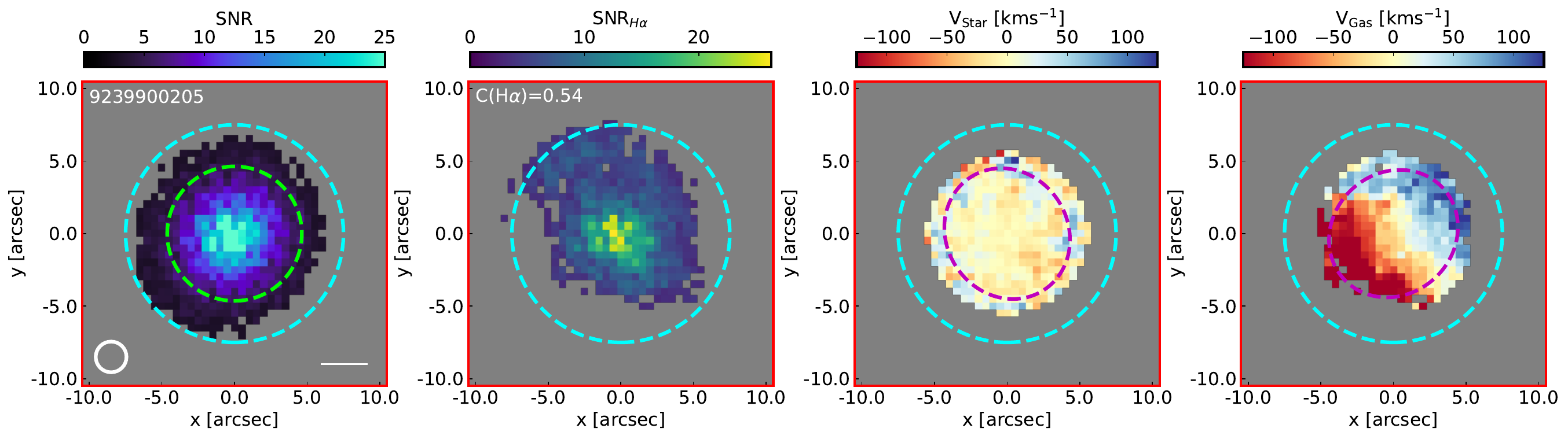}
    \caption{The same as Fig. \ref{9008500107} but for galaxy 9239900205}
    \label{9239900182}
\end{figure*}

\begin{figure*}
    \centering
    \includegraphics[width=1\linewidth]{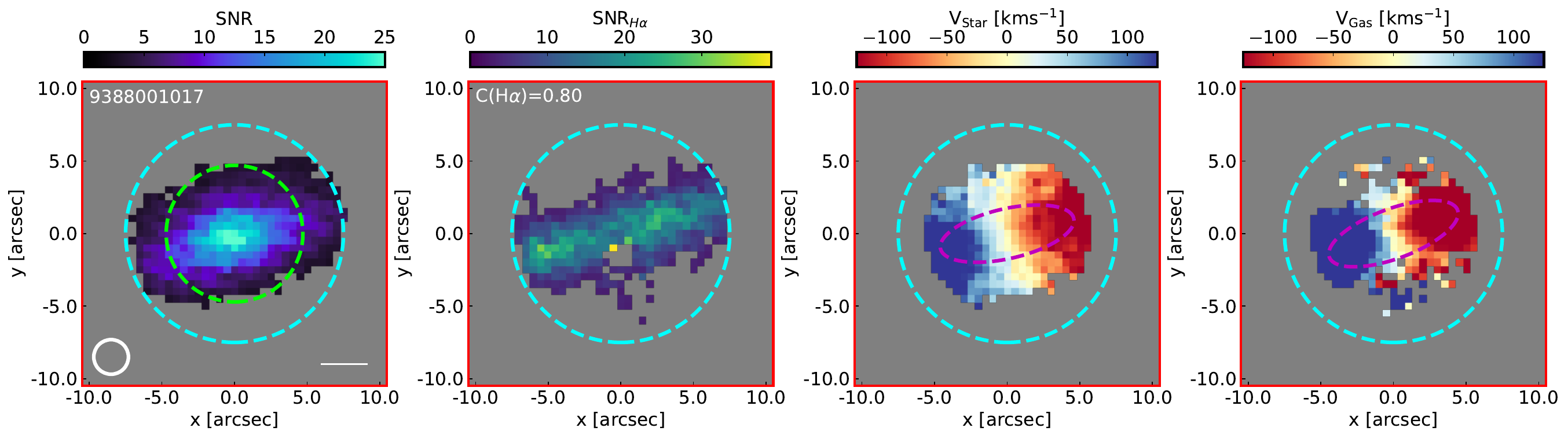}
    \caption{The same as Fig. \ref{9008500107} but for galaxy 9388001017}
    \label{9239901328}
\end{figure*}

\begin{figure*}
    \centering
    \includegraphics[width=1\linewidth]{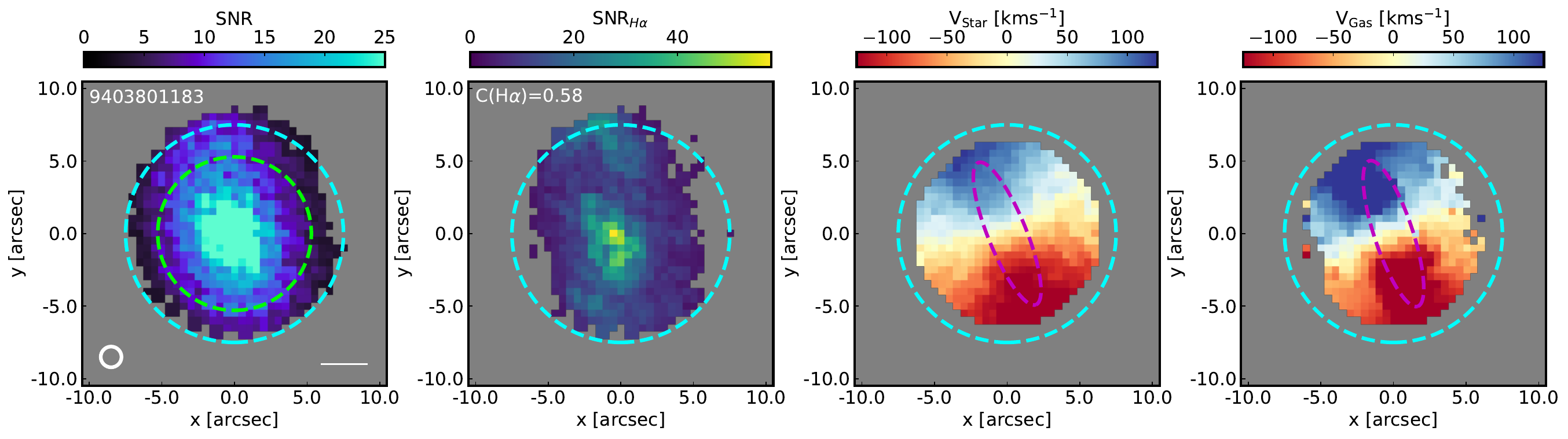}
    \caption{The same as Fig. \ref{9008500107} but for galaxy 9403801183}
    \label{9239901328}
\end{figure*}


\bsp	
\label{lastpage}
\end{document}